\documentclass[useAMS,usenatbib]{mnras}
\usepackage{journals}
\usepackage{longtable}
\usepackage{pdflscape}
\usepackage{graphicx}

\title[Supernovae and their host galaxies -- IV]{Supernovae and their host galaxies -- IV.
The distribution of supernovae relative to spiral arms}
\author[L.~S.~Aramyan~et~al.]{L.~S.~Aramyan,$^{1}$\thanks{E-mail: aramyan@bao.sci.am}
A.~A.~Hakobyan,$^{1}$\thanks{E-mail: hakobyan@bao.sci.am}
A.~R.~Petrosian,$^{1}$
V.~de~Lapparent,$^{2}$
\newauthor
E.~Bertin,$^{2}$
G.~A.~Mamon,$^{2}$
D.~Kunth,$^{2}$
T.~A.~Nazaryan,$^{1}$
V.~Adibekyan$^{3}$
\newauthor
and M.~Turatto$^{4}$
\\
$^{1}$Byurakan Astrophysical Observatory, 0213 Byurakan, Aragatsotn province, Armenia\\
$^{2}$Institut d'Astrophysique de Paris (UMR 7095: CNRS \& UPMC), 98bis Bd Arago, F-75014 Paris, France\\
$^{3}$Instituto de Astrof\'{\i}ísica e Ci\^{e}ncia do Espa\c{c}o, Universidade do Porto, CAUP, Rua das Estrelas, P-4150-762 Porto, Portugal\\
$^{4}$INAF -- Osservatorio Astronomico di Padova, Vicolo dell'Osservatorio 5, I-35122 Padova, Italy}
\begin{document}

\date{Accepted ... . Received ... ; in original form ...}

\pagerange{\pageref{firstpage}--\pageref{lastpage}} \pubyear{2016}

\maketitle

\label{firstpage}

\begin{abstract}

	Using a sample of 215 supernovae (SNe), we analyze their
	positions relative to the spiral arms of their host galaxies,
	distinguishing grand-design (GD) spirals from non-GD (NGD) galaxies.
	We find that: (1) in GD galaxies, an offset exists
	between the positions of Ia and core-collapse (CC) SNe
	relative to the peaks of arms, while in NGD galaxies
	the positions show no such shifts;
	(2) in GD galaxies, the positions of CC SNe
	relative to the peaks of arms are correlated
	with the radial distance from the galaxy nucleus.
	Inside (outside) the corotation radius, CC SNe
	are found closer to the inner (outer) edge.
  No such  correlation is observed for SNe in NGD galaxies
	nor for SNe Ia in either galaxy class;
	(3) in GD galaxies, SNe~Ibc occur closer to the
	leading edges of the arms than do
	SNe~II, while in NGD galaxies they are more concentrated towards
	the peaks of arms.
	In both samples of hosts, the distributions of SNe Ia
	relative to the arms have broader wings.
	These observations suggest that shocks in spiral arms of GD galaxies
	trigger star formation in the leading edges of arms
	affecting the distributions of CC SNe (known to have
	short-lived progenitors). The closer locations of SNe~Ibc vs. SNe~II
  relative to the leading edges of the arms supports the belief
	that SNe~Ibc have more massive progenitors. SNe~Ia having less massive
	and older progenitors, have more time to drift away
	from the leading edge of the spiral arms.
	
\end{abstract}


\section{Introduction}

Many studies have been performed to find the links between
different types of supernovae (SNe) and stellar populations of host galaxies
(e.g. \citealt{2001MNRAS.328.1181N,2009MNRAS.399..559A,2009A&A...508.1259H};
\citealt*{2012MNRAS.424.2841H};
\citealt{2012ApJ...759..107K,2012A&A...542A..29G,
2013Ap&SS.347..365N,2015MNRAS.448..732A}),
and estimate the possible ages of SN progenitors
\citep[e.g.][]{2012MNRAS.424.1372A,2013AJ....146...30K,
2013AJ....146...31K,2013MNRAS.428.1927C}.
Studies of the spatial distribution of SNe in host galaxies
give further insights on the nature of their progenitors.
In particular, the relations between distributions of SNe and star formation in spiral arms are
important in this context (e.g. \citealt*{1994PASP..106.1276B};
\citealt{1996ApJ...473..707M,2005AJ....129.1369P};
\citealt*{2007AstL...33..715M}).

It is well known that star forming regions in spiral discs are generally concentrated
in spiral arms \citep[e.g.][]{2002MNRAS.337.1113S}.
There are a variety of known structures of spiral galaxies,
with different numbers and shapes of
their arms \citep[for recent review see][]{2013pss6.book....1B}.
According to their spiral features, spiral galaxies are divided into three main types:
1) grand-design (GD) spirals with typically two arms;
2) flocculent spirals with many short arms;
and 3) multi-armed spirals with several long spirals.
Spiral arms in GD galaxies are thought to be density waves and may in fact represent quasi-steady
wave modes \citep[e.g.][]{1989ApJ...338...78B,1996ApJ...457..125Z,1998ApJ...499...93Z,1999ApJ...518..613Z}.
However, there is also another interpretation, largely based on simulations, in which such arms are
short-lived and recurrent structures \citep[see review by][]{2013pss5.book..923S}.
GD galaxies are usually associated with stellar bars
\citep[e.g.][]{1976ApJ...209...53S,1989ApJ...342..677E,2013JKAS...46..141A},
or density waves caused by the tidal field of a nearby neighbor
(e.g. \citealt{1972ApJ...178..623T,1979ApJ...233..539K};
\citealt*{2011MNRAS.414..538K}; \citealt{2013MNRAS.429.1051C}).
In contrast to GD spiral galaxies, flocculent and multi-armed
spiral galaxies are likely formed from gravitational instabilities,
or are sheared star formation regions (e.g. \citealt{1982FCPh....7..241S}; \citealt*{2003ApJ...590..271E}).

The distribution of stellar ages in spiral arms
have been studied in GD galaxies.
Investigating the dynamics of spiral galaxies,
\citet{1969ApJ...158..123R} proposed that the piled up gas
in a spiral arm experiences a strong shock
that triggers star formation.
He suggested that in the case of trailing spiral arms,
young stars and HII regions inside the corotation radius are located in the inner edges of
observable arms, and in the case of leading spirals they occur in the outer edges.
Many studies were performed to find
the predicted offsets between tracers of star formation
(e.g. \citealt{1988Natur.334..402V}; \citealt*{1993A&A...274..123G})
using eye-determination of the offsets,
which could potentially affect the results.
Applying a numerical method on 14 nearby disc galaxies,
\citet{2008AJ....136.2872T} found that the radial dependence of the azimuthal offset
between the HI and $24~\mu m$ emission is in agreement with the prediction of \citet{1969ApJ...158..123R}.
However, using the same method for another sample,
\citet{2011ApJ...735..101F} found no systematic ordering of angular offsets.

The study of different types of SNe as tracers of star formation
caused by density waves and their distributions relative to spiral arms
can help to better constrain the nature of SN progenitors.
Types Ib, Ic, and II SNe, collectively called core-collapse (CC) SNe,
are considered to be a result of gravitational collapse of young massive stellar cores
in the final stage of the stellar evolution with initial masses
${\geq 8~M_\odot}$.
Moreover, CC SNe are well associated with star-forming sites
\citep[e.g.][]{1996AJ....111.2017V,2001AstL...27..411T,2006A&A...453...57J,
2008MNRAS.390.1527A,2012MNRAS.424.1372A}.
SNe~Ia are thought to originate from a thermonuclear explosion of C/O white dwarfs
\citep[e.g.][]{2012PASA...29..447M}.
The rate of SNe~Ia in spiral galaxies suggests
that a significant fraction of their progenitors
belongs to the stellar population with intermediate ages
and they can be considered as weak tracers of
star formation (e.g. \citealt*{1999A&A...351..459C};
\citealt{2005A&A...433..807M,2011MNRAS.412.1473L,2011Ap.....54..301H}).

In the pioneering study of \citet{1973PASP...85..564M},
the locations of 19 SNe in host galaxies were investigated.
He found that over half of the SNe occur at the innermost edge of a spiral arm,
and estimated that their lifetimes are less than five million years
with masses greater than $35~M_\odot$.
\citet{1976ApJ...204..519M} measured the positions of 84 SNe relative to the spiral arms and
found that the locations of SNe~II are more concentrated relative to the arms
than are the locations of SNe~I
(at that time SN types Ia, Ib, and Ic were not considered separately).
Using a larger sample, \citet{1994PASP..106.1276B} studied the distribution of
SNe~Ia, Ib/c, and II relative to spiral arms and found that all SNe are distributed closer
to the spiral arms comparing with random distribution in discs.
\citet{1996ApJ...473..707M} and \citet{2005AJ....129.1369P} found
that CC SNe are more tightly concentrated to the arms than Ia.
Finally, \citet{2007AstL...33..715M} found that
SNe~Ibc\footnote{{\footnotesize By SN Ibc, we denote stripped-envelope SNe
of Type Ib, Ic, and mixed Ib/c whose specific subclassification is uncertain.}}
are concentrated to the inner edges of the spiral arms.
They noted that there is a difference between
the distribution of Ibc and II SNe, and conclude that the progenitors of Ibc SNe are younger.

These studies used different samples of SNe,
various methods of analysis and inhomogeneous data of SNe and their host galaxies.
In particular, the spiral arms were traced by images of different quality,
which is caused by various distances of the objects, exposure times, filters, etc.
In addition, widths, shapes, and edges of spiral arms usually were determined visually,
which also could affect the results.
\defcitealias{2012A&A...544A..81H}{Paper~I}
\defcitealias{2014MNRAS.444.2428H}{Paper~II}
\defcitealias{2016MNRAS.456.2848H}{Paper~III}

In our first paper of this series
\citep[][hereafter Paper~I]{2012A&A...544A..81H} we have reported the creation of a large and
well-defined data base that combines extensive new measurements and
a literature search of 3876 SNe and their 3679 host galaxies
located in the sky area covered by the
SDSS Data Release 8 (DR8).
This data base is much larger than all previous ones, and
provides a homogenous set of global parameters of SN hosts, including morphological
classifications and measures of activity classes of nuclei.
In addition, we have analysed and discussed many selection effects and biases,
which usually affect the studies of SNe.
In the second paper \citep[][hereafter Paper~II]{2014MNRAS.444.2428H},
we have analysed the number ratios of different SN types in spirals
with various morphologies and in barred and unbarred galaxies,
different levels of morphological disturbance, and activity classes of nucleus.
We proposed that the underlying mechanisms shaping the number ratios
of SN types could be interpreted within the framework
of interaction-induced star formation, in addition
to the known relations between morphologies
and stellar populations.
In the third paper \citep[][hereafter Paper III]{2016MNRAS.456.2848H},
we have presented an analysis of the impact of bars and bulges
on the radial distributions of SNe in the stellar discs of S0--Sm host galaxies.
We suggested that the additional mechanism shaping the distributions
of Type Ia and CC SNe can be explained within the framework
of substantial suppression of massive star formation in the radial range
swept by strong bars, particularly in early-type spirals.
We refer the reader to 
Papers~I, II and III
for more details.

In this fourth paper of the series, we investigate the distribution of different types of SNe
relative to spiral arms taking into account the intrinsic properties of arms in order to
find links between the distributions of the various SN types and arm's stellar populations.
Moreover, considering possible differences of the distributions of
various stellar populations in GD
and multi-armed/flocculent spirals \citep[e.g.][]{2010MNRAS.409..396D,2011MNRAS.415..753S},
we investigate the distribution of SNe in these subsamples of host galaxies.

The outline of this paper is as follows: Section~\ref{samplesel} describes the sample.
In Section~\ref{datared}, we provide a description 
of the spiral arm classification and extraction from the
images, and the arm tracing method in the vicinity of the SN.
We present our results in Section~\ref{Resdiscus} and discuss them
in Section~\ref{discus}.
Our conclusions are summarized in Section~\ref{concl}.
Throughout this paper, we adopt a flat cosmological model with
$\Omega_{\rm m}=0.27$, $\Omega_{\rm \Lambda}=0.73$ and a Hubble constant
$H_0=73 \,\rm km \,s^{-1} \,Mpc^{-1}$ \citep{2007ApJS..170..377S},
both to conform with the values used in our data base \citepalias{2012A&A...544A..81H}.

\section{Sample selection}
\label{samplesel}

The sample of this study is drawn from the catalog
of \citetalias{2012A&A...544A..81H},\footnote{Parameters
of several SNe and their host galaxies were revised in \citet{2013Ap.....56..153A}.}
which contains 3876 SNe (72 SNe~I, 1990 SNe~Ia, 234 SNe~Ibc, 870 SNe~II, and 710 unclassified SNe)
from the area covered by the SDSS DR8 and homogeneously measured parameters of their host galaxies.
Note that for SNe~II it includes also subtypes IIP, IIL, IIn, and IIb.
The last SN included in the sample of \citetalias{2012A&A...544A..81H}
is SN~2011bl, discovered on 2011 April~5.

The analysis of the distribution of SNe relative to spiral arms requires a well-defined sample
and high angular resolution images of the SNe hosts.
We selected our sample of SNe and their hosts according to the following criteria:

\begin{itemize}
  \item SNe with available spectroscopic classification;
  \item SNe with available measurement of position\footnote{The SN position is usually provided by the SN
	catalogues via its offset from the host galaxy nucleus, or equatorial coordinates.};
  \item SNe located outside the circumnuclear region (we set a lower limit
  of 15 per cent of the host galaxy angular radius
	$R_{\rm SN}/R_{25}~>~0.15$)\footnote{$R_{25}$ is the host galaxy
	\textit{g}-band 25th magnitude isophotal radius $R_{25} = D_{25}/2$.},
	as well as outside any bar, interarm HII
	region\footnote{In bulge+disc subtracted images
	SNe 2001ed (Ia) and 1961I (II) are located
	on interarm HII regions and thus are excluded from our sample.}
	or overlapping region of two spiral arms;
  \item host galaxies with Sa--Sd morphological types;
  \item host galaxies without a strong morphological disturbance
	(excluding interacting, merging and post-merging classes
	of \citetalias{2014MNRAS.444.2428H});
  \item host galaxies with angular size $D_{\rm 25} \geq 40''$ (to better distinguish the spiral arms);
  \item host galaxies with inclination $i \le 40^\circ$ (to minimize absorption and projection effects);
  \item host galaxies whose surface brightness are not strongly contaminated
	by foreground Milky Way stars\footnote{We selected only host galaxies for which at least two
	of inner edge, outer edge, and the arm peak were determined.}.

\end{itemize}

After applying these restrictions,
the final sample of our study consists of 215 SNe in 187 host galaxies.
The distribution of the different types of SNe according to the host galaxies morphological
types is given in Table~\ref{sample}.

\begin{table}
  \centering
  \begin{minipage}{69mm}
  \caption{Distribution of SN types according to
  the morphological classification of their host galaxies}
  \label{sample}
  \tabcolsep 5pt
  \begin{tabular}{lrrrrrrrr}
  \hline
  &\multicolumn{1}{c}{Sa}
  &\multicolumn{1}{c}{Sab}&\multicolumn{1}{c}{Sb}&\multicolumn{1}{c}{Sbc}&\multicolumn{1}{c}{Sc}&\multicolumn{1}{c}{Scd}
  &\multicolumn{1}{c}{Sd}&\multicolumn{1}{r}{All}\\
  \hline
    Ia & 5 & 4 & 8 & 16 & 25 & 5 & 2 & 65 \\
    Ibc & 2 & 1 & 1 & 9 & 9 & 2 & 1 & 25 \\
    II & 0 & 1 & 19 & 21 & 58 & 13 & 13 & 125 \\
  \hline
    All & 7 & 6 & 28 & 46 & 92 & 20 & 16 & 215 \\
  \hline \\
  \end{tabular}
  \parbox{\hsize}{SNe~Ib and Ic are not considered separately for lack of
    sufficient numbers (only 4 SNe~Ib).
Among SNe~II, 6 are of Type IIb and 18 are of Type IIn.}
  \end{minipage}
\end{table}

\section{Data reduction}
\label{datared}

In our study, the background subtracted and photometrically calibrated
$g$-band images from the SDSS DR8 were used:
among the SDSS channels with good signal-to-noise ratio ($g$, $r$, $i$),
the arm-interarm contrast is the highest in the $g$-band, as it
traces the young stellar populations in the spiral arms.
These images were further analyzed to distinguish the spiral arms
from the interarm regions. The whole reduction process is given below.

\subsection{Determination of spiral arm classes}
\label{Armclassdet}

The spiral arm classes of all 187 host galaxies were determined
visually from the $g$-band SDSS images, following the spiral arm 
classification of \citet{1987ApJ...314....3E}, who gave a numerical
designation of arms according to flocculence and degree of chaos.
The galaxies with arm class 12 contain only two long and symmetric arms dominating the optical disc,
and the ones with class 9 have two inner symmetric arms, and multiple long outer arms.
In all other cases the arms are fragmented (classes between 1 and 4),
feathery/irregular (classes between 5 and 7) or tightly warped/ring-like (class 8).
Classes 10 and 11 were previously denoted to bars
and companions and are no longer used.
Then the galaxies were assigned as GD (classes 9 and 12) or
non-GD (NGD, all classes except 9 and 12).

According to \citet{1987ApJ...314....3E}, the most common mis-classifications of arm classes are
from 2 or 3 to 4 or 5 (or vice versa) and from 9 to 12 (or vice versa). Because we separate
the host galaxies arm classes into two broad subsamples GD
and NGD, the possible misclassification of GD into NGD (or vice versa) is negligible.

In order to test our classification,
the whole sample of host galaxies was classified twice by two different coauthors.
By comparing these two classifications, we determined
that our classification was 97 per cent reliable.

\subsection{Spiral arm structures}

\begin{figure}
\centering
  \includegraphics[width=0.49\hsize]{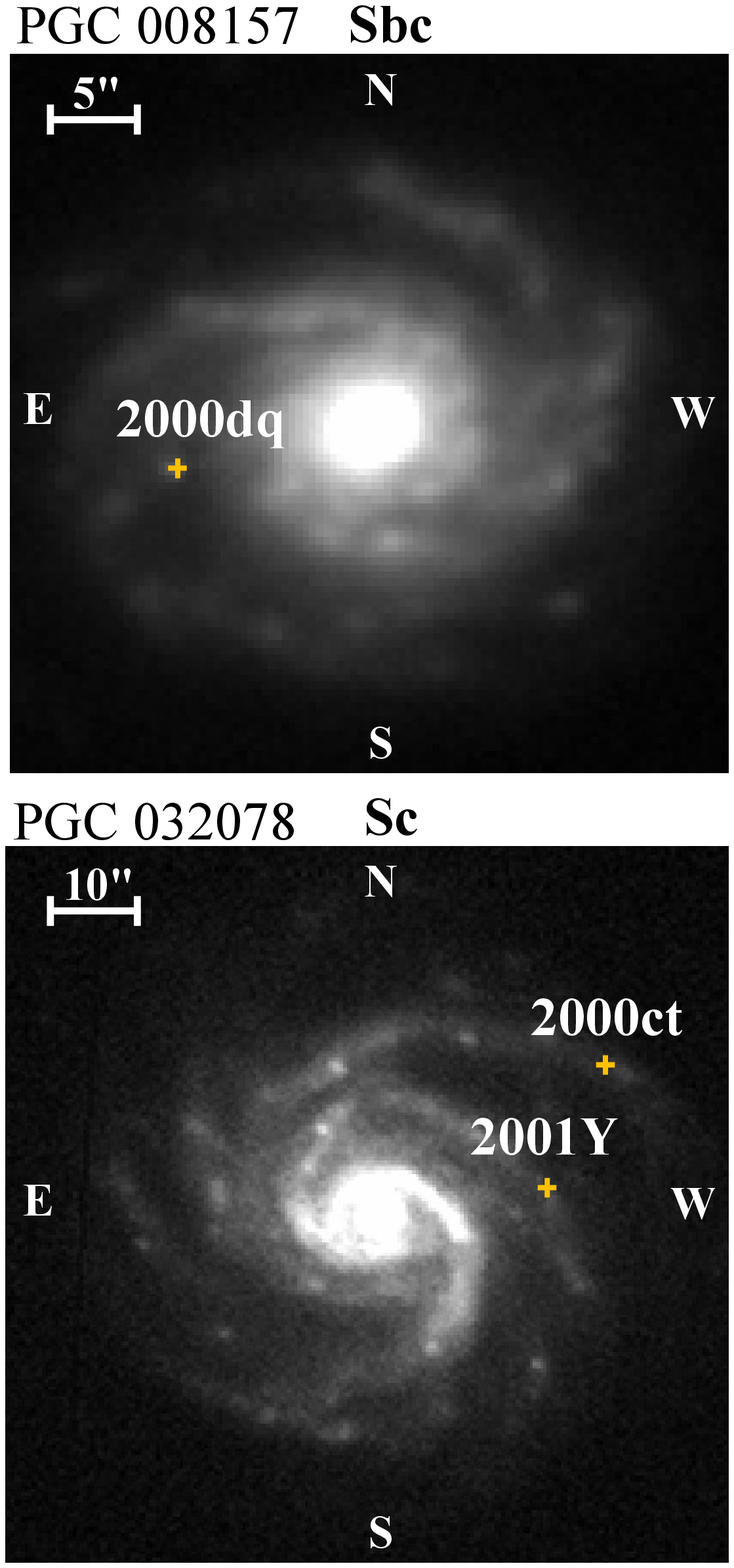} 
  \includegraphics[width=0.49\hsize]{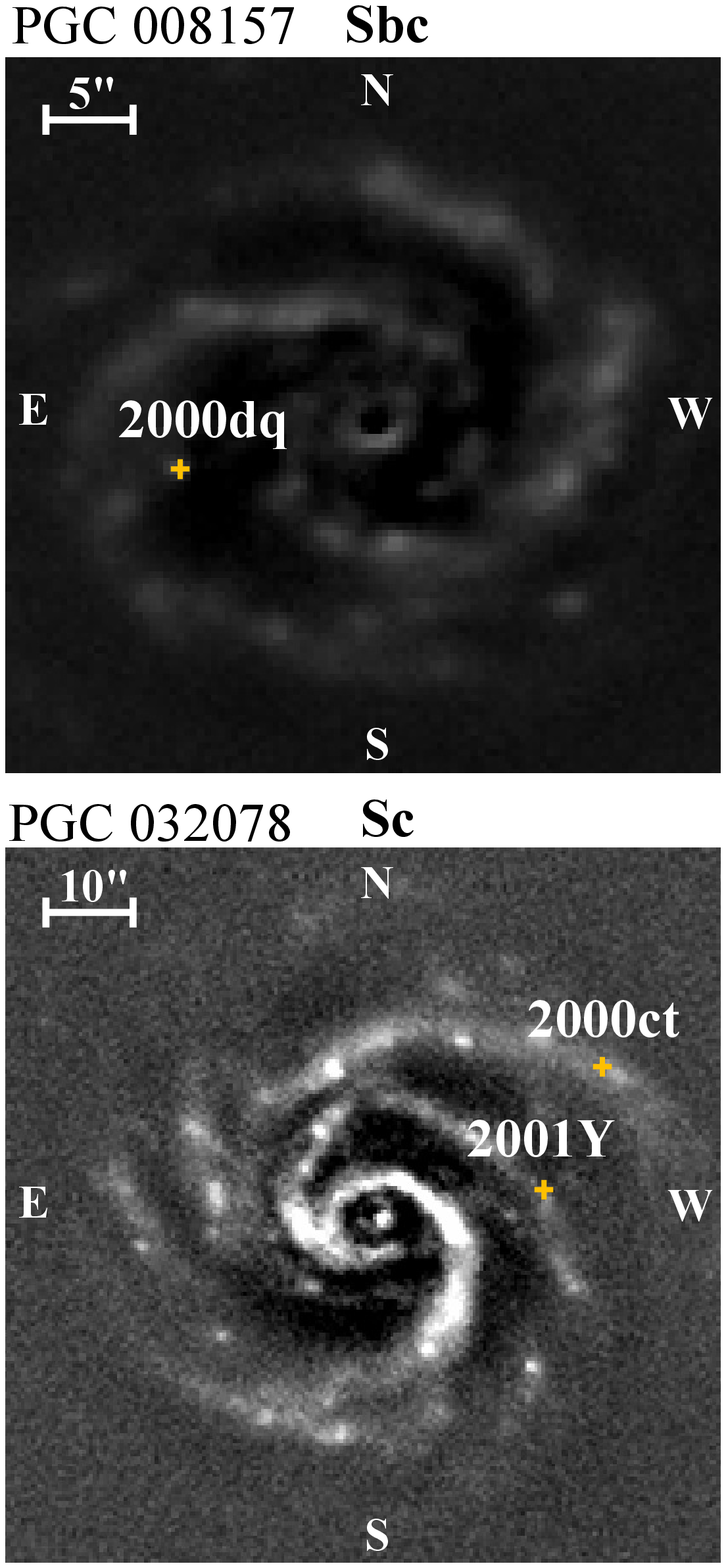}
  \caption{Examples of the original (left) SDSS $g$-band and
	         bulge+disc subtracted (right) images of SNe host galaxies.
           The PGC object identifiers and morphological types are listed at the top.
           The SN names and positions (marked by a cross sign) are also shown.
           In all images, north is up and east to the left.}
  \label{resids}
\end{figure}

Using the version~2.18.4 of {\sc SExtractor}
software \citep{1996A&AS..117..393B},
we carried out a procedure to isolate the spiral structure of the galaxies.
We first fitted all $g$-band SDSS images of the host galaxies 
in the sample with bulge+disc models ($r^{1/4}$ bulge
and exponential disc profiles are used for all galaxies).

\begin{figure}
  \centering
  \includegraphics[width=\hsize]{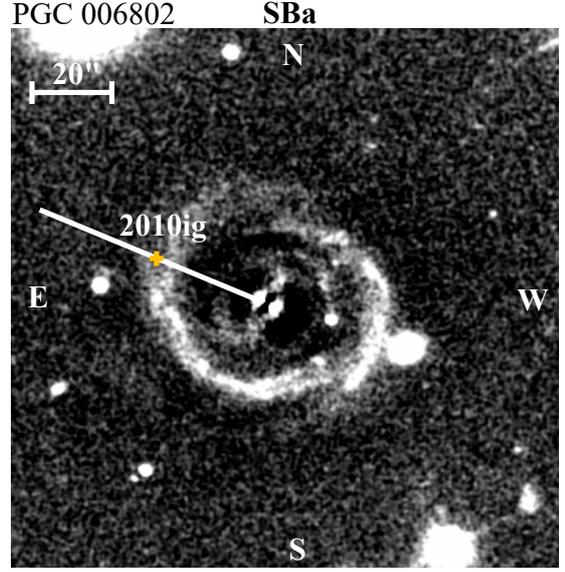} 
  \includegraphics[width=\hsize]{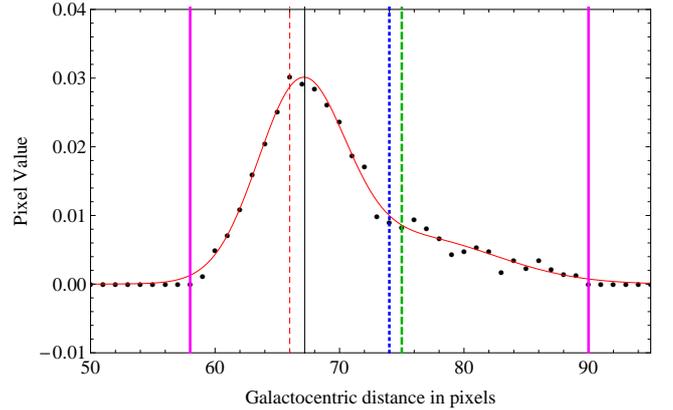}
  \caption{\emph{Upper panel:} residual SDSS $g$-band image of the host galaxy of SN 2010ig,
	         after subtracting the best fit bulge+disc components.
           The PGC object identifier and morphological type are listed at the top of the figure
           (north is up and east to the left).
           The SN name, position (marked by a cross sign), and SN radial line are also shown.
           \emph{Bottom panel:} brightness profile along the SN radial line and its fit (thin red line)
					 with a multi-peak Gaussian.
           The vertical lines represent the position of the SN (green thick dashed),
					 the edges (thick magenta), middle (blue thick dotted)
					 and brightest pixel (thin dashed red),
					 and peak of Gaussian fit (thin black) of the spiral arm.}
  \label{profile}
\end{figure}

The modeled bulge+disc was then subtracted from each original image.
In the following, we describe how we use the residual image
for determining the location of the spiral arms
and their radial profile relative to the position of the SN.
Fig.~\ref{resids} illustrates the procedure on two spiral galaxies.

\subsection{Position of SN with respect to spiral arm}
\label{Armposition}

\begin{figure*}
\centering
	\includegraphics[width=0.44\hsize]{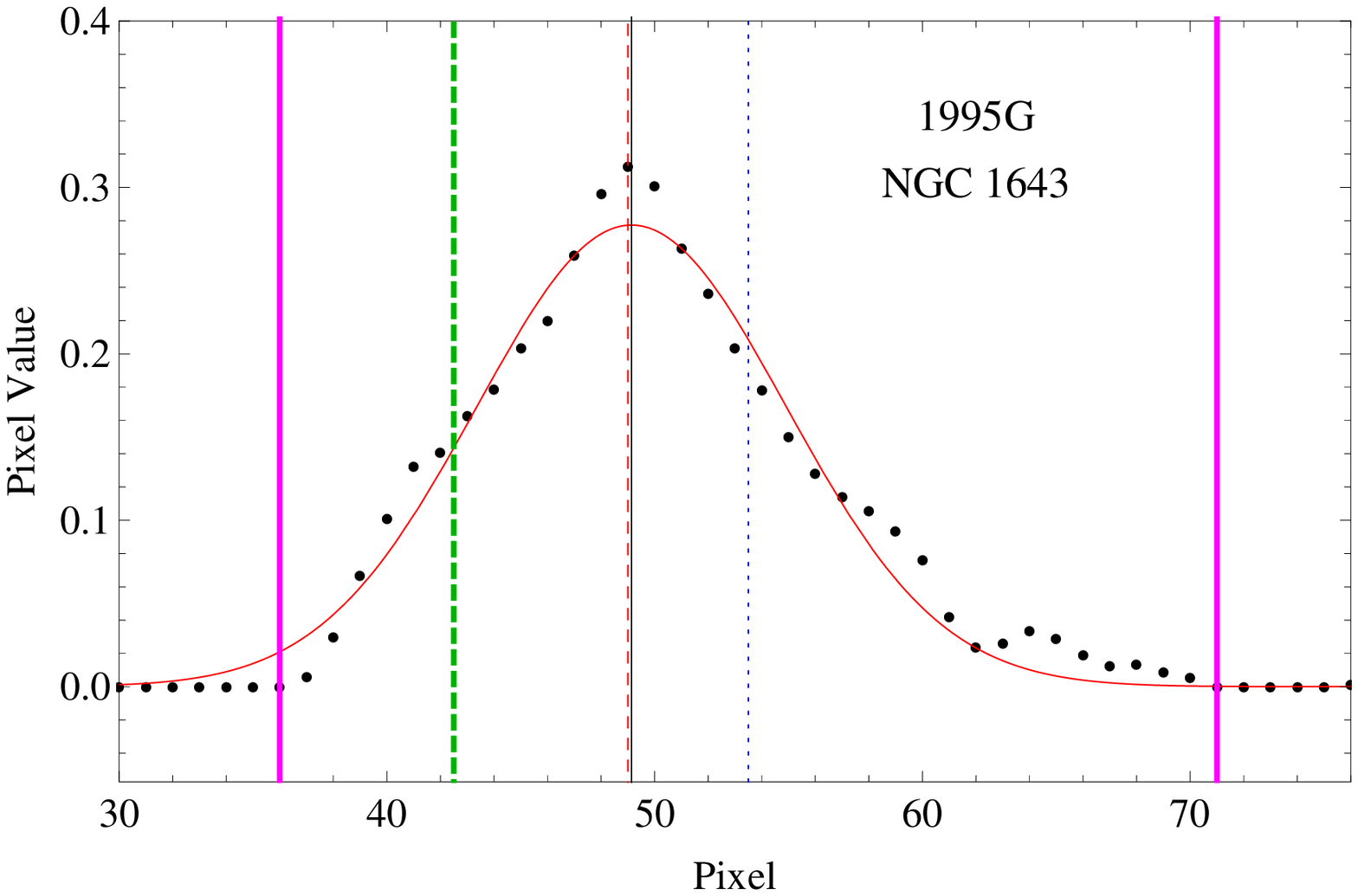} 
	\includegraphics[width=0.44\hsize]{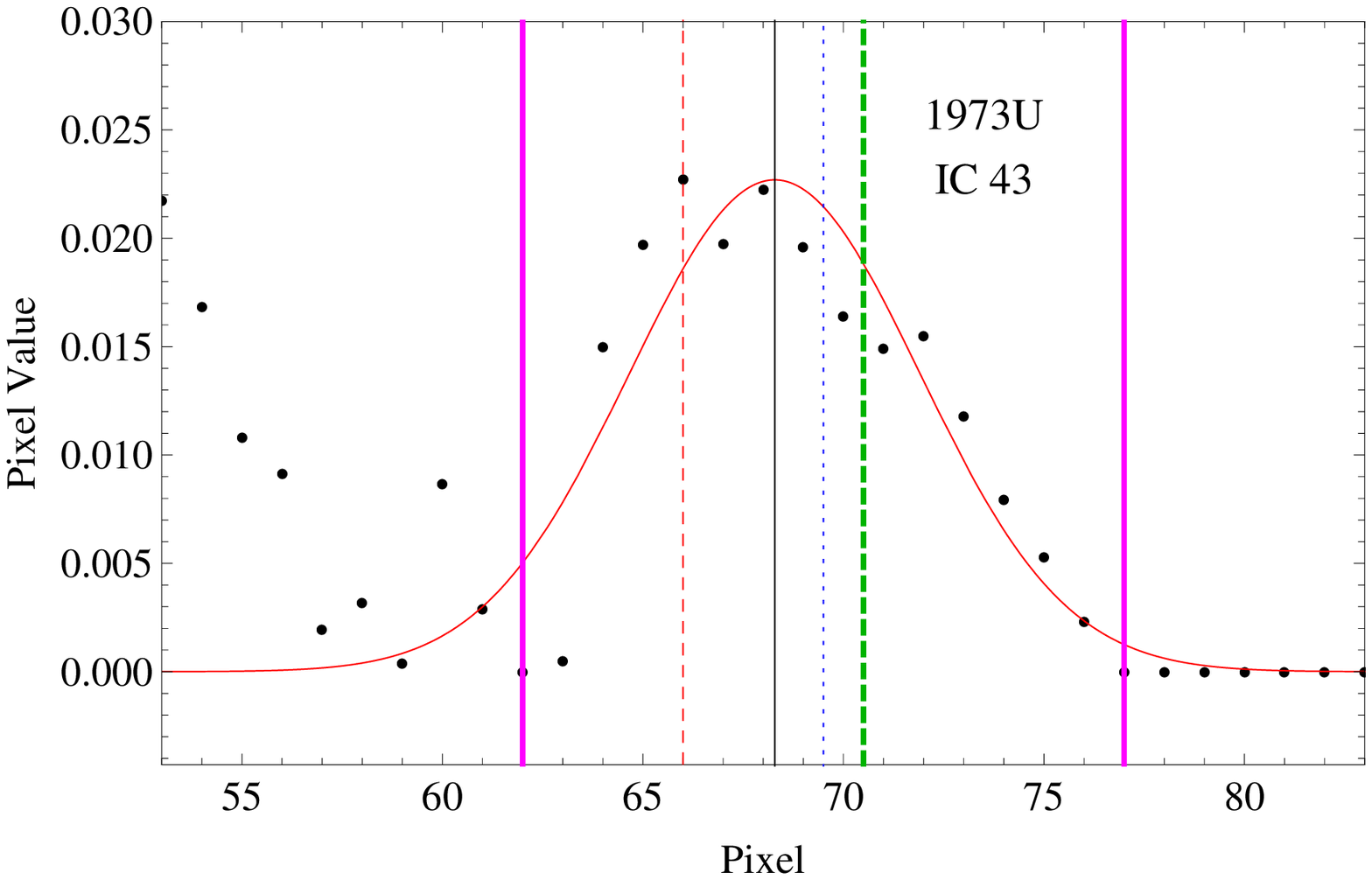} \\
	\includegraphics[width=0.44\hsize]{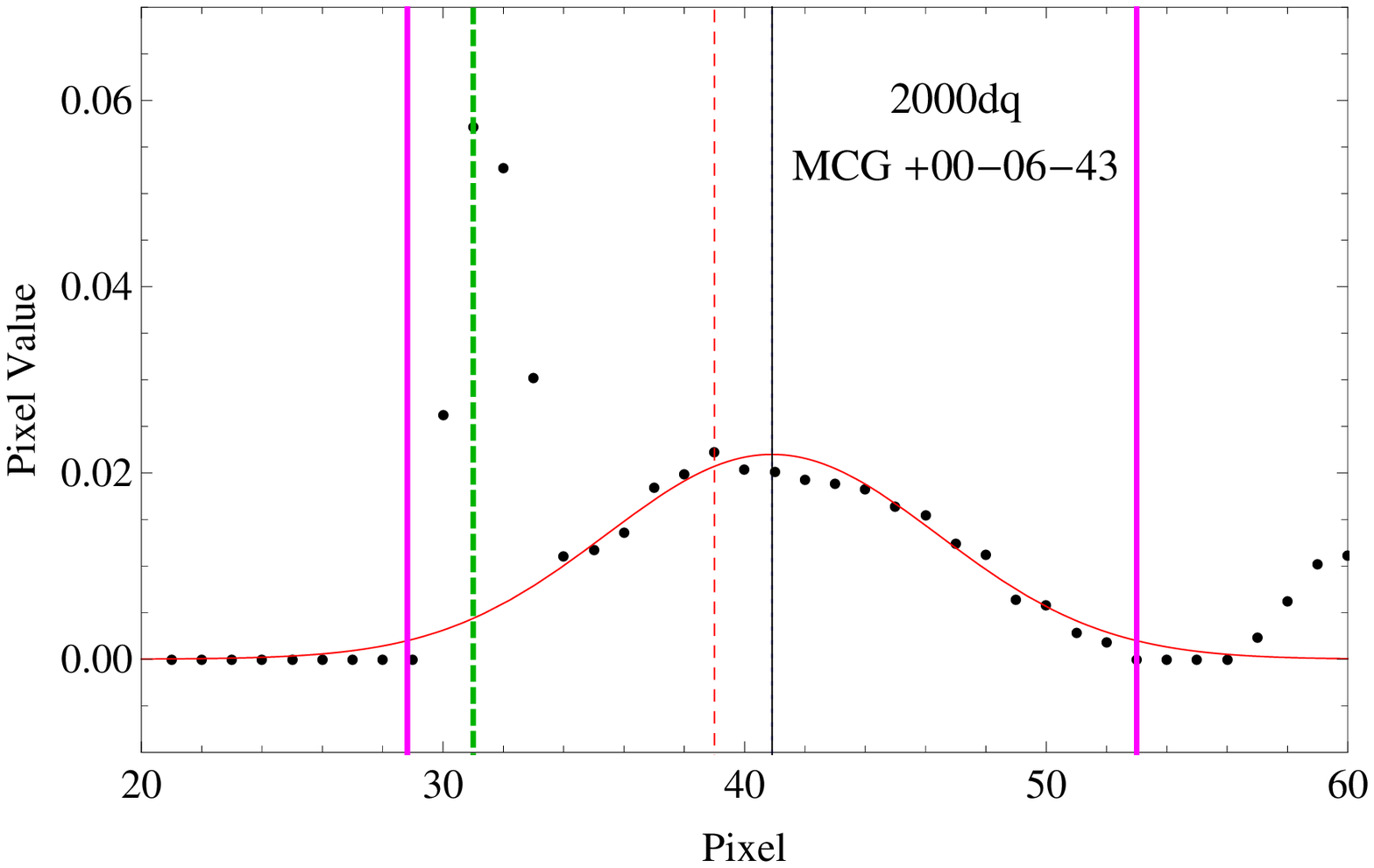} 
	\includegraphics[width=0.44\hsize]{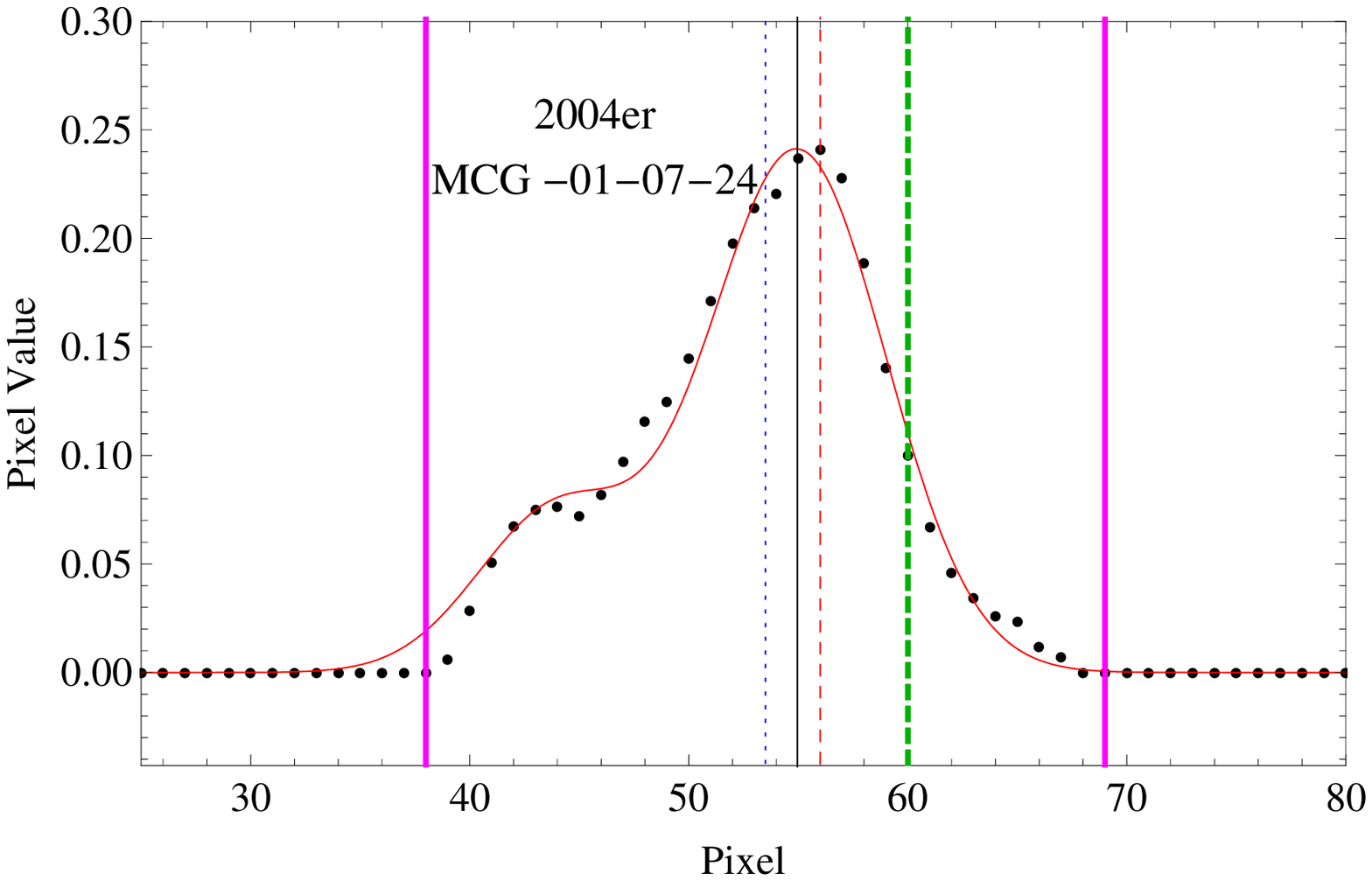} \\
	\includegraphics[width=0.44\hsize]{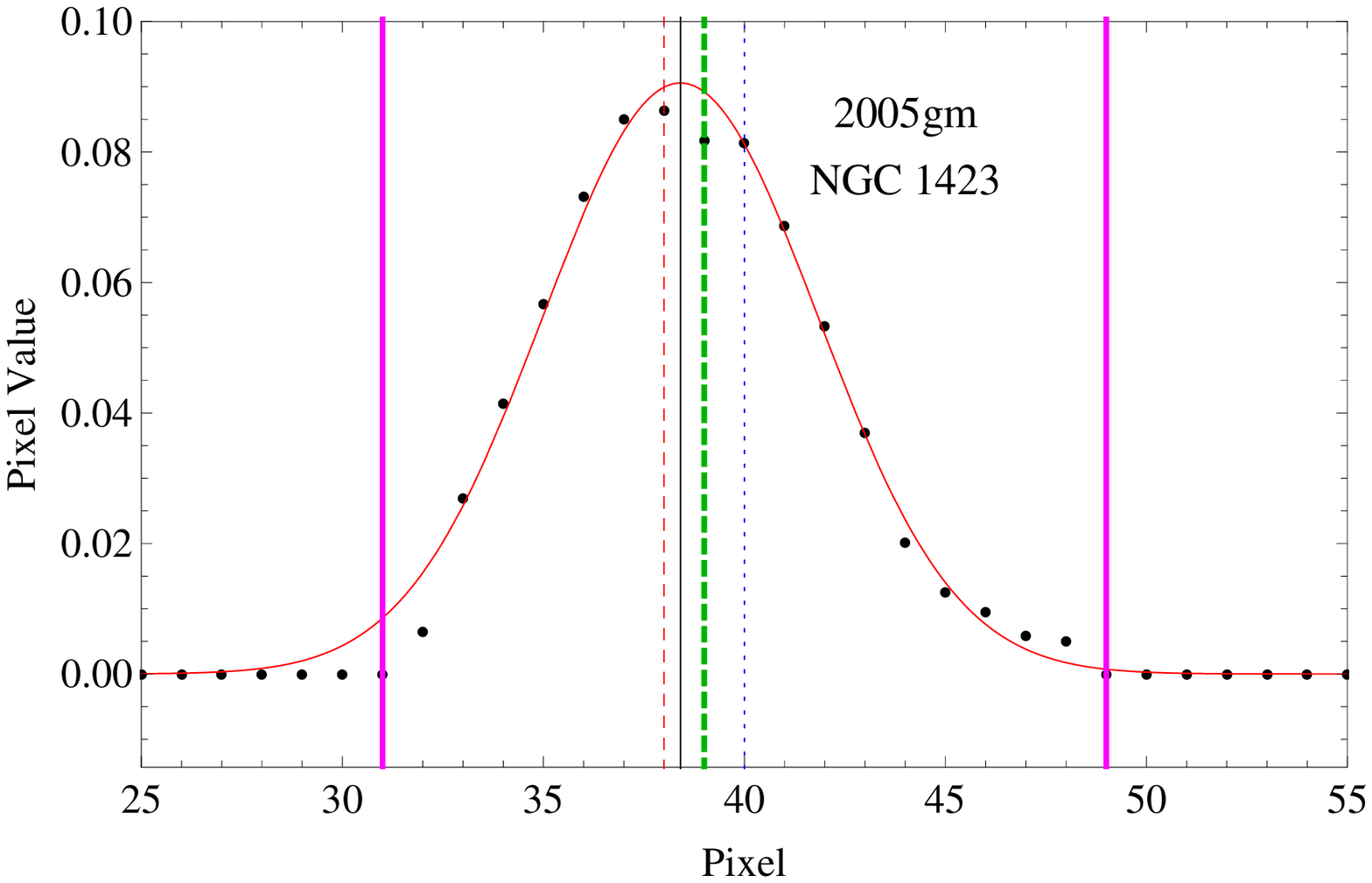} 
	\includegraphics[width=0.44\hsize]{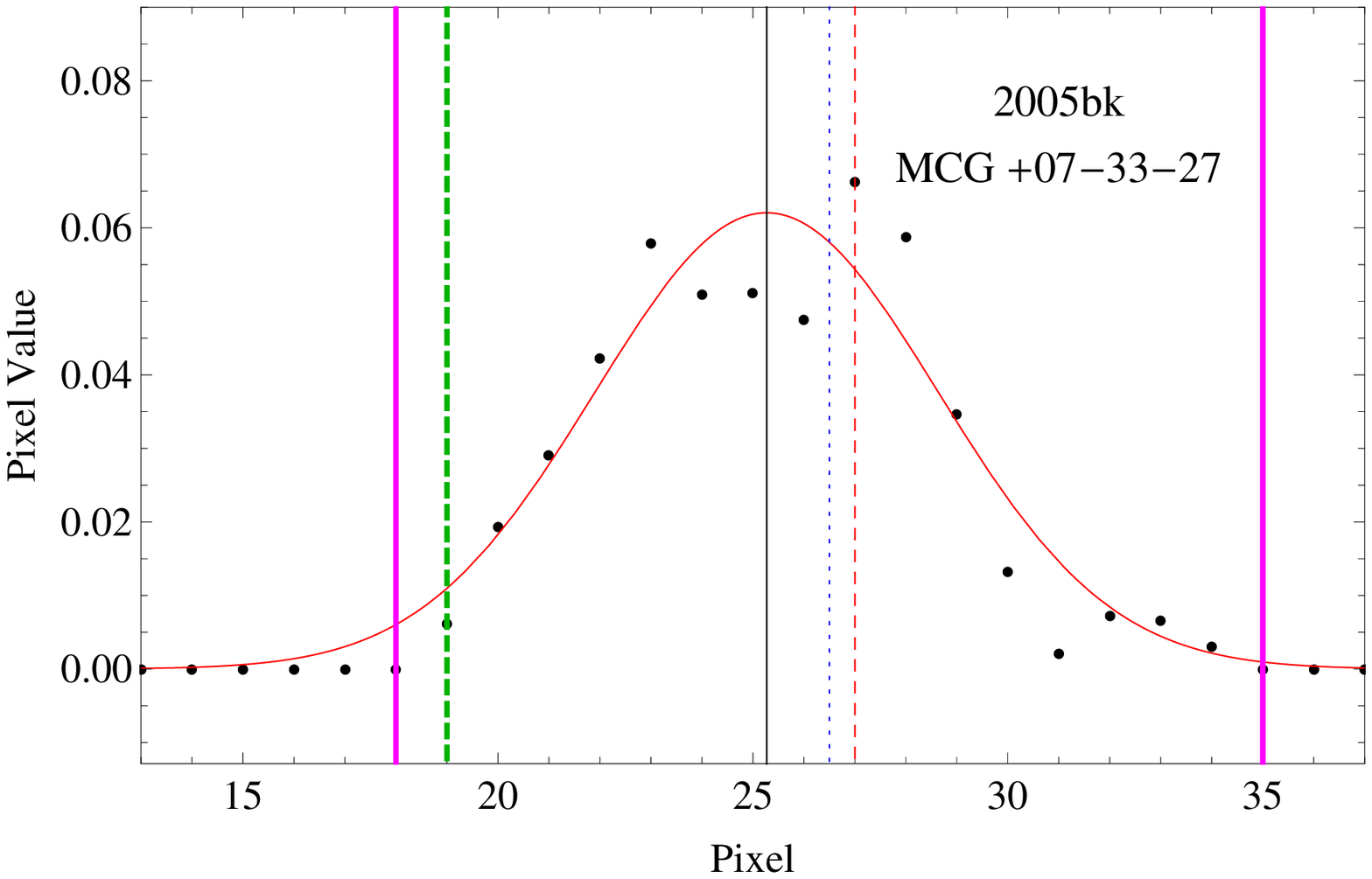} \\
	\includegraphics[width=0.44\hsize]{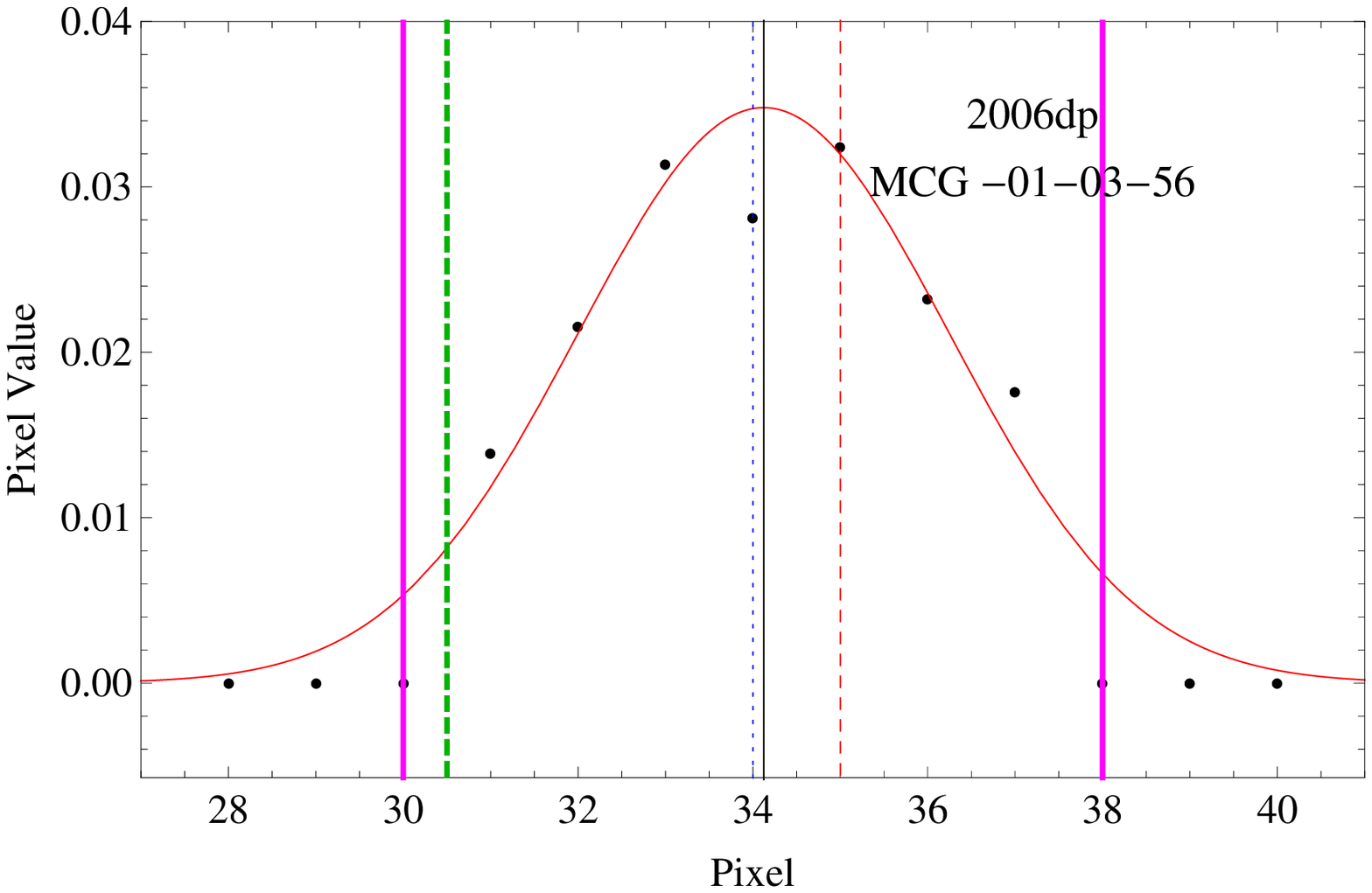} 
	\includegraphics[width=0.44\hsize]{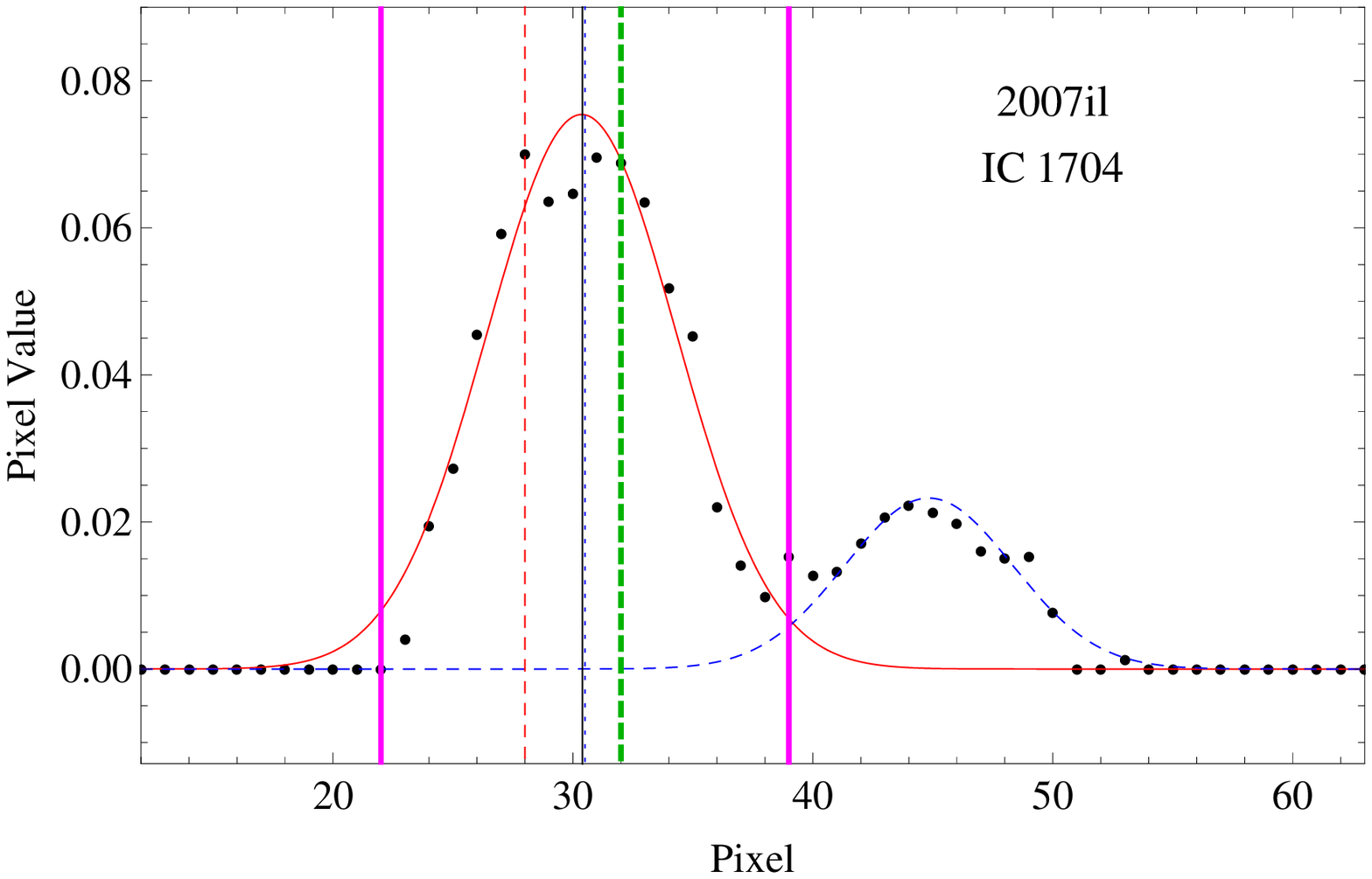}
  \caption{Four representative spiral arm intensity profiles
	        for GD (\emph{left}) and NGD (\emph{right}) host galaxies.
					The SN positions, inner and outer edges, middles, as well
					the peaks and brightest pixels of spiral arms are shown with vertical lines,
					similar to Fig.~\ref{profile}. For SN~2007il, the best-fit profile
					of the contamination of the neighbor arm is shown as the dashed blue curve.}
  \label{GDNGDprofiles}
\end{figure*}

To avoid effects due to different distances of the host galaxies,
the image of the host galaxy is smoothed with a Gaussian filter
with $\sigma = 0.25~{\rm pixel} \times 100/D$,
where $D$ is the comoving distance of the galaxy in Mpc.
Because the exponential fits of the discs use light
from both the arm and interarm regions, the values of pixels in the interarm regions of
the residual images are negative. These are replaced with a null value,
and a zero threshold defines the inner and outer borders of the spiral arms.

To study the distribution of SNe with respect to spiral arms,
at first the closest spiral arm structure to the SN is defined
(for \textit{interarm} SN we define the closest spiral arm
as the one with the closest edge).
The closest segment of spiral arm to SN position is fixed along the radial line passing
through the galaxy nucleus and SN location (e.g. upper panel of Fig.~\ref{profile}).
The radial light profile is obtained using the SAO DS9 analysis tool
with the \textit{linear region} option.
The coordinates of the nucleus and
the isophotal radius $R_{\rm 25}$
of the host galaxies are those provided
in the data base of \citetalias{2012A&A...544A..81H}.
Accordingly,  arm profiles are constructed (bottom panel of Fig.~\ref{profile}).
Four representative arm profiles of GD spirals and four arm profiles
of NGD spirals are given respectively in the left- and right-hand panels of
Fig.~\ref{GDNGDprofiles}.

We define \emph{arm} and \emph{interarm} SNe, those that respectively lie
inside the arm borders or  in the interarm region, respectively.

In the cases when the profile is contaminated by star(s) or by the SN
(when the SN is still visible in the SDSS image,
e.g. SN~2000dq in Figs.~\ref{resids} and \ref{GDNGDprofiles}),
we fit the contaminant with a Gaussian profile and
subsequently subtract it from the global profile.
In the cases when the edges of several spiral arms overlap,
we fit the profile with a multi-Gaussian function
(the number of components equals the number of visible spiral arms).
We also apply  a multi-Gaussian when the brightest pixel is shifted
from the middle of arm borders by more than 10 per cent of the arm width,
or when the profile of the arm shows a multi-peak structure.

\subsection{Normalized distances along arm profiles}
\label{normcor}

To study the SNe distributions relative to spiral arms, two different approaches
for the normalization of the distances are applied.
In the first approach, the distance\footnote{{\footnotesize For all radial distance measurements, we
used the deprojection method described in \citet{2009A&A...508.1259H}.}}
$d_{\rm SN}$ of an SN
from the inner edge (at galactocentric radius $r=r_{-}$) of the spiral arm
is normalized 
to the width $W$, defined as the distance between inner and outer edge (at
galactocentric radius $r_+$) of the arm
\citep[e.g.][]{1976ApJ...204..519M,1994PASP..106.1276B,2007AstL...33..715M}:
\begin{equation}
d_1 =  {d_{\rm SN}\over W}= {r_{\rm SN} - r_{-}\over r_+-r_{-}}  \ ,
\label{d1}
\end{equation}
where $W = r_{+}-r_{-}$ is the width of the arm along the radius vector
passing through the galaxy nucleus, as well as $r_{-}$ and $r_{+}$ are the radial distances
of inner and outer edges, respectively.
It is important to note that due to
the orbital motion of disc and spiral arms, the
$r_{\rm SN} - r_{-}$ distance depends on
the pitch angle of spiral arm, that can vary from galaxy to galaxy.
However, the same dependence is true also for the arm width ($r_+~-~r_{-}$),
determined through the same radial direction.
Therefore, the normalized distance has the advantage of being independent
from the pitch angle of the considered segment of spiral
arm.\footnote{It is worth mentioning that we consider
only the orbital motion of SN progenitors,
neglecting the possible radial migration. In Sect.~\ref{discus},
we discuss in more detail the motion of progenitors in the discs.}
According to this definition, \textit{interarm} SNe close to the inner edge
of the spiral arm have 
$d_1 < 0$, while \textit{interarm} SNe
closer to the outer edge have 
$d_1 > 1$.

We  test
whether the location of the peak of the best-fit multi-Gaussian
changes according to the choice of the threshold
value defining the inner and outer borders of the spiral arms.
To this end, 10 arm profiles were randomly selected
and for each 12 different thresholds were applied.
These thresholds were uniformly distributed
between 2/3 of the peak value and $1.5\,\sigma$
above the mean interarm value.
For each threshold, we fit the arm profiles and determine the positions of the peaks.
The test shows that the standard deviations of the distributions
of the positions of arm peaks are about $0.02-0.03$ of the widths of arms.

\begin{figure}
  \begin{center}
  \includegraphics[width=1\hsize]{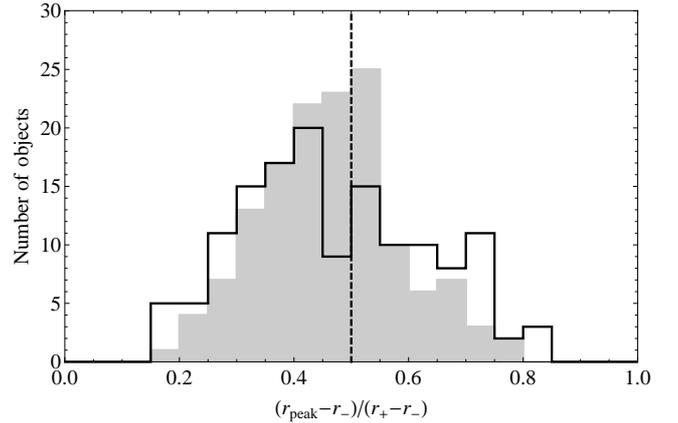}
  \end{center}
  \caption{Distribution of the brightest pixels along spiral arms (open histogram)
	         and peaks of Gaussian fits (shaded histogram) in the $g$-band images.
					 The abscissa is zero and 1 at the inner
					 and outer edges of the spiral arms, respectively.	
	         The cases when at least one of the edges is contaminated and
           not well-determined, are excluded from the histogram.
           The dashed line shows the middle of the arm.}
  \label{gpeakdist}
\end{figure}

Then, we pay attention to the distributions of $g$-band Gaussian peaks
and brightest points in the arm profiles.
Fig.~\ref{gpeakdist} shows that 
the normalized distribution of both the brightest points and the peaks
of the profiles are generally located
closer to the inner edge than to the outer edge
of the spiral arms.
In other words, the arm profiles tend to be positively skewed.

In order to check the normality,
we compare the distributions of peaks and brightest points in spiral arms
with the Gaussian functions with fixed $\mu = 0.5$
(the mean of the distributions) and free $\mu$, respectively.
For this purpose we perform a maximum likelihood estimation to find
the standard deviations and $\mu$ (for the second case) of the Gaussian.
We then compare the distribution of
$(r_{\rm peak}-r_{\rm -})/(r_{\rm +}-r_{\rm -})$ distances
with the best fit functions.
To examine the significance of the differences
between the distributions in different samples
(or between the sample and a best fit function),
we employ the Kolmogorov-Smirnov (KS) and Anderson-Darling (AD) tests\footnote{The AD test
is similar to the KS test, except that it is
more sensitive to differences in the tails of distributions.}.
The corresponding KS and AD probabilities,
$P_{\mu=1/2}^{\rm KS}$ and $P_{\mu=1/2}^{\rm AD}$,
are given in Cols.~3 and 4, as well as 
$P_{{\rm free}-\mu}^{\rm KS}$ and $P_{{\rm free}-\mu}^{\rm AD}$
which are given in Cols.~5 and 6 of Table~\ref{peakstatsnorm}.
Table~\ref{peakstatsnorm} shows that both distributions
are Gaussians that are significantly shifted from the middle of spiral arms
towards the inner edges.

\begin{table}
  \centering
  \caption{The mean, standard deviation, $P_{\rm KS}$ and $P_{\rm AD}$ values
					 of the normality tests of the $(r_{\rm peak}-r_{\rm -})/(r_{\rm +}-r_{\rm -})$ distances
					 of arm peaks and brightest points (BP). The statistically significant differences are
					 highlighted in bold.}
  \label{peakstatsnorm}
  \tabcolsep 2.5pt
  \begin{tabular}{lrrcccc}
    \hline
     & \multicolumn{1}{c}{$\mu$} & \multicolumn{1}{c}{$\sigma$} &
		\multicolumn{1}{c}{$P_{\mu=1/2}^{\rm KS}$} & \multicolumn{1}{c}{$P_{\mu=1/2}^{\rm AD}$} &
		\multicolumn{1}{c}{$P_{{\rm free}-\mu}^{\rm KS}$} & \multicolumn{1}{c}{$P_{{\rm free}-\mu}^{\rm AD}$} \\
		& \multicolumn{1}{c}{(1)} & \multicolumn{1}{c}{(2)} &
		\multicolumn{1}{c}{(3)} & \multicolumn{1}{c}{(4)} & \multicolumn{1}{c}{(5)} &
		\multicolumn{1}{c}{(6)} \\
		\hline
 Peak & $0.47 \pm 0.01$ & $0.12 \pm 0.01$ & \textbf{0.001} & \textbf{0.001} & 0.803 & 0.884 \\
 BP & $0.47 \pm 0.01$ & $0.16 \pm 0.01$ & \textbf{0.003} & \textbf{0.005} & 0.329 & 0.321 \\
\hline 
  \end{tabular}
\end{table}

\begin{figure}
  \begin{center}$
  \begin{array}{cc}
  \includegraphics[width=1\hsize]{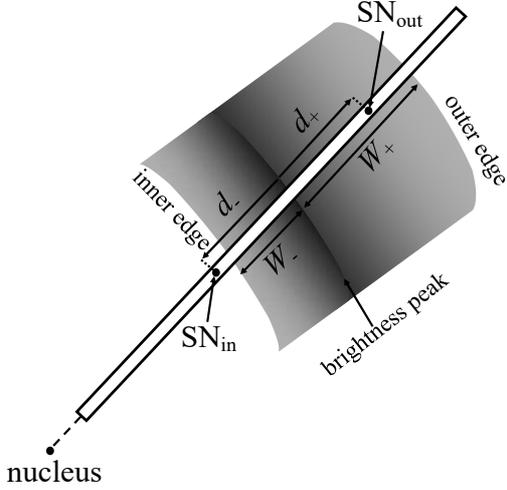}
  \end{array}$
  \end{center}
  \caption{The scheme of normalization according to the location of the SN relative to spiral arm:
           when the SN is located in the inner part of the spiral arm (${\rm SN_{in}}$), and
           when the SN is located in the outer part of the arm (${\rm SN_{out}}$).}
  \label{normaliz}
\end{figure}

It is important to note that the brightest points
of the arm profiles are often associated with local HII regions and
do not represent the general distributions
of the $g$-band light profiles of the spiral arms.
We therefore design a second normalization
using either the distance between the peak
of the spiral arm and its inner edge ($W_{-} = r_{\rm peak}-r_{-}$),
or that between the peak and the outer edge ($W_+ = r_+-r_{\rm peak}$),
depending on whether the SN is located 
between the host galaxy nucleus and the
peak of the spiral arm or outside this region (labeled
${\rm SN_{in}}$ and ${\rm SN_{out}}$ respectively in Fig.~\ref{normaliz}).
The new normalized distance is then
\begin{equation}
d_2 = {d_\pm \over W_\pm }
=  {r_{\rm SN}-r_{\rm peak} \over 
\displaystyle |r_{{\rm sign}(r_{\rm
      SN}-r_{\rm peak})}-r_{\rm peak}|} \ ,
\label{d2}
\end{equation}
where 
\[
r_{\rm sign}(r_{\rm SN}-r_{\rm peak})= \left \{
\begin{array}{ll}
-  & \hbox { for } r_{\rm SN} < r_{\rm peak} \ , \\
+  & \hbox { for } r_{\rm SN} > r_{\rm peak} \ .
\end{array}
\right.
\]

It is clear that, if we normalize the distances of SNe relative to spiral arms
to the widths of arms, the distribution of normalized distances will be systematically
shifted and enhanced to the inner part of spiral arms with respect to the second case.
Most probably this effect is reflected in the results of \citet{2007AstL...33..715M},
who found that the distribution of SNe~Ibc is skewed towards the inner edges
of arms.

We apply this normalization both when the SN is located inside
the borders of the spiral arm (e.g. $\rm SN_{out}$ in Fig.~\ref{normaliz}),
and when the SN is located outside the borders of spiral arm
(e.g. $\rm SN_{in}$ in Fig.~\ref{normaliz}).

\section{Results}
\label{Resdiscus}

In this section, we present a detailed analysis of the distributions of SNe with different locations
in GD and NGD spiral host galaxies.
In particular, the distributions of SNe in the arm and interarm regions
and their dependences on galactocentric distances are analyzed.
Among the 215 SNe, 178 are arm SNe, 
while 37 are interarm SNe.
The full list of arm and interarm SNe
(Col.~1), their types (Col.~2),
host galaxies names (Col.~3), morphological types (Col.~4),
spiral arm classes (Col.~5), the SNe galactocentric distances (Col.~6),
normalized distances $d_1$ (Col.~7) and $d_2$ (Col.~8) are given in
Tables~\ref{arm} and \ref{interarm}.

\subsection{Distribution of SNe relative to spiral arms}
\label{insidearm}

The comparison of the distributions of different types of SNe located in spiral arms and
interarm regions (Tables~\ref{arm} and \ref{interarm}) shows that SNe~Ibc are discovered
only within spiral arms. The majority ($88 \pm 2$ per cent) of SNe~II also belongs to
spiral arms, while this percentage is only $66^{+3}_{-4}$ per cent
for SNe~Ia\footnote{The errors on fractions are calculated
using the Bayesian approach of \citet{2011PASA...28..128C}.}.

\begin{figure}
\centering
  \includegraphics[width=\hsize]{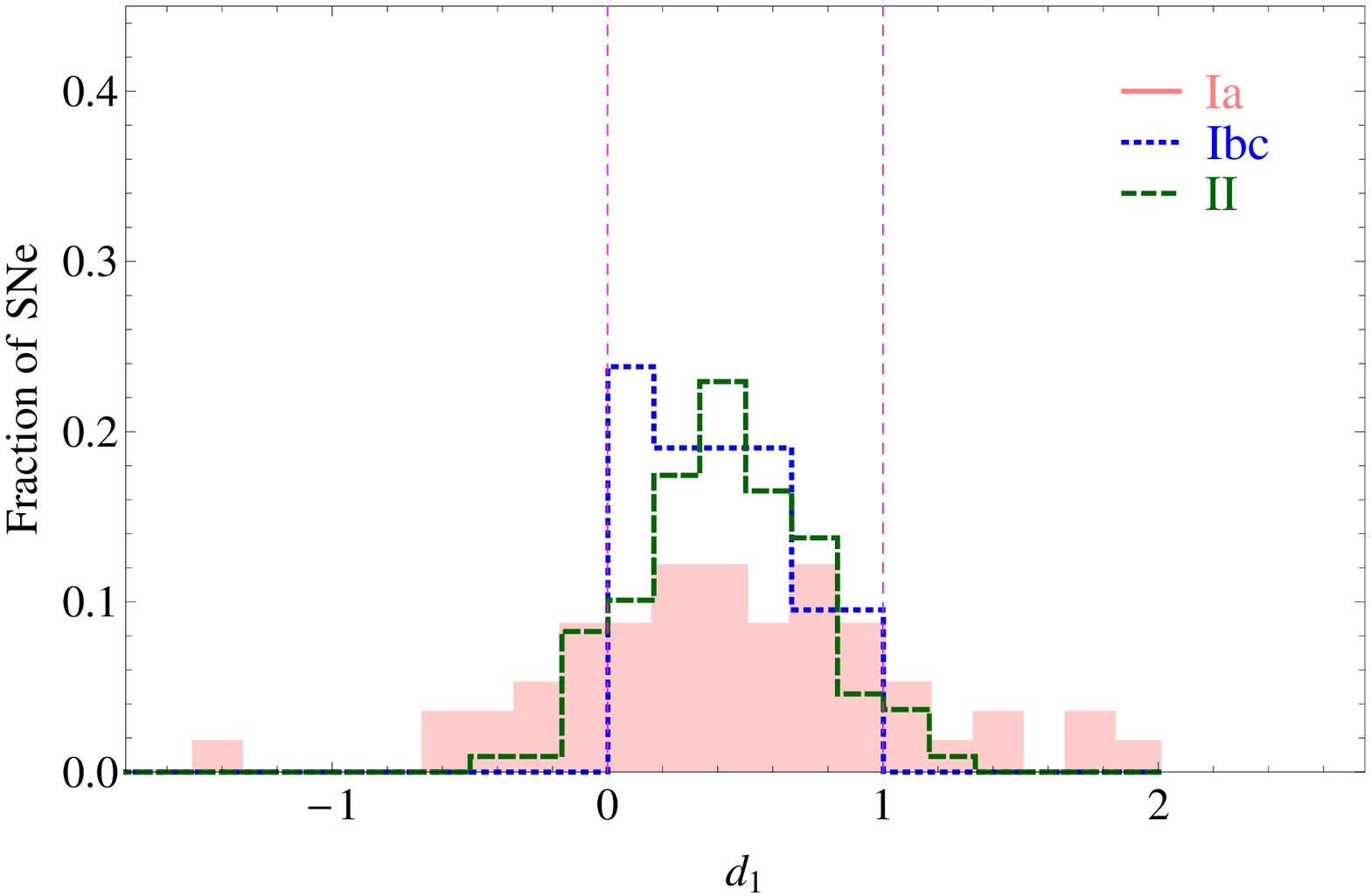}

\bigskip

  \includegraphics[width=\hsize]{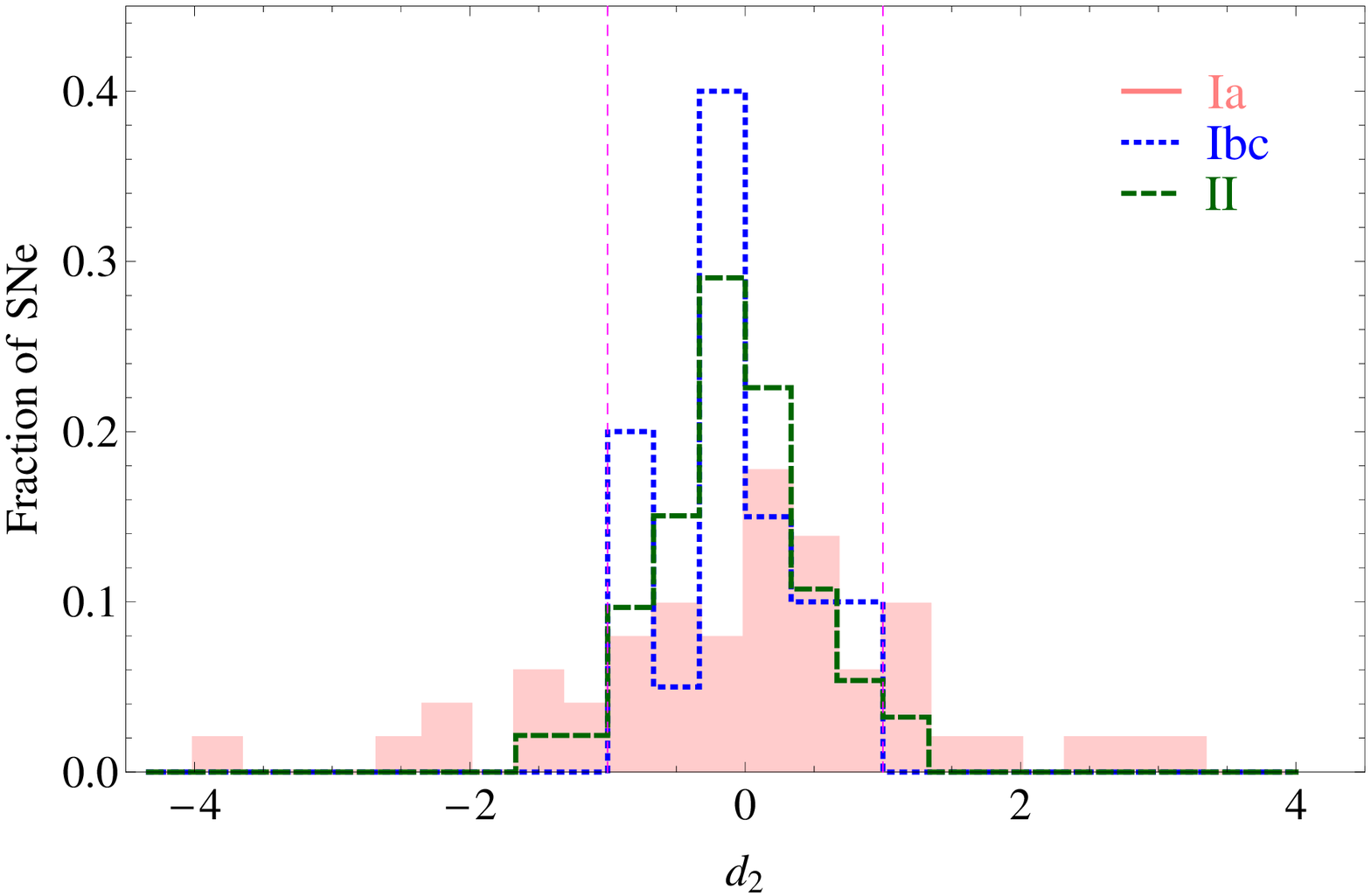}
\caption{\emph{Upper panel:} distributions of SN positions relative to the
				 inner edge of the spiral arm normalized to the width of the arm
				 (see eq.~[\ref{d1}]).
         \emph{Bottom panel:} distributions of
				 SNe positions relative to the peaks of arms
				 normalized to the inner/outer semi-widths
				 (see eq.~[\ref{d2}]).
         The data flagged by `*' symbols in Tables~\ref{arm} and \ref{interarm}
				 are not considered in both panels.
         The filled red histograms represent SNe~Ia, while the
				 green dashed and blue dotted histograms
				 represent SNe~II and Ibc, respectively.}
  \label{distr1}
\end{figure}

\begin{table}
  \centering
  \caption{The mean, standard deviation, $P_{\rm KS}$ and $P_{\rm AD}$ values
					 of the normality tests of the $d_1$ distances.
					 The statistically significant difference is highlighted in bold.}
  \label{d1statsnorm}
  \tabcolsep 2.5pt
  \begin{tabular}{lrrcccc}
    \hline
    SN & \multicolumn{1}{c}{$\left\langle d_1 \right\rangle$} & \multicolumn{1}{c}{$\sigma$} &
		\multicolumn{1}{c}{$P_{\mu=1/2}^{\rm KS}$} & \multicolumn{1}{c}{$P_{\mu=1/2}^{\rm AD}$} &
		\multicolumn{1}{c}{$P_{{\rm free}-\mu}^{\rm KS}$} & \multicolumn{1}{c}{$P_{{\rm free}-\mu}^{\rm AD}$} \\
		& \multicolumn{1}{c}{(1)} & \multicolumn{1}{c}{(2)} &
		\multicolumn{1}{c}{(3)} & \multicolumn{1}{c}{(4)} & \multicolumn{1}{c}{(5)} &
		\multicolumn{1}{c}{(6)} \\
		\hline
 Ia & $0.46 \pm 0.08$ & $0.63 \pm 0.07$ & 0.967 & 0.931 & 1.000 & 0.997 \\
Ibc & $0.42 \pm 0.05$ & $0.26 \pm 0.03$ & 0.280 & 0.209 & 0.807 & 0.848 \\
 II & $0.43 \pm 0.03$ & $0.32 \pm 0.02$ & 0.097 & \textbf{0.039} & 0.977 &
 0.999 \\
\hline 
  \end{tabular}
\end{table}

\begin{table}
  \centering
  \caption{The mean, standard deviation, $P_{\rm KS}$ and $P_{\rm AD}$ values
					 of the normality tests of the $d_2$ distances.}
  \label{d2statsnorm}
  \tabcolsep 2pt
  \begin{tabular}{lrrcccc}
    \hline
    SN &
		\multicolumn{1}{c}{$\left\langle d_2 \right\rangle$} & \multicolumn{1}{c}{$\sigma$} &
		\multicolumn{1}{c}{$P_{\mu=0}^{\rm KS}$} & \multicolumn{1}{c}{$P_{\mu=0}^{\rm AD}$} &
		\multicolumn{1}{c}{$P_{{\rm free}-\mu}^{\rm KS}$} & \multicolumn{1}{c}{$P_{{\rm free}-\mu}^{\rm AD}$}\\
		& \multicolumn{1}{c}{(1)} & \multicolumn{1}{c}{(2)} &
		\multicolumn{1}{c}{(3)} & \multicolumn{1}{c}{(4)} & \multicolumn{1}{c}{(5)} &
		\multicolumn{1}{c}{(6)} \\
		\hline
   Ia &
	 $0.01 \pm 0.18$ & $1.31 \pm 0.16$ & 0.815 & 0.733 & 0.850 & 0.743 \\
	 Ibc &
	 $-0.07 \pm 0.10$ & $0.46 \pm 0.06$ & 0.443 & 0.615 & 0.818 & 0.823 \\
	 II & 
	 $-0.10 \pm 0.06$ & $0.56 \pm 0.04$ & 0.116 & 0.162 & 0.880 & 0.917 \\
		\hline 
  \end{tabular}
\end{table}

The distributions of \textit{arm} and \textit{interarm} SNe, with different
normalizations discussed in Sect.~\ref{normcor}, are presented in Fig.~\ref{distr1}.
The mean and standard deviation of the distributions of $d_1$
and $d_2$, with their bootstrapped ($10^{4}$ times) errors are given in
Tables~\ref{d1statsnorm} and \ref{d2statsnorm}, respectively.
The upper panel of Fig.~\ref{distr1} shows the distributions
of SNe distances relative to the inner edges of spiral arms
normalized to the width of the arm.
This figure and Table~\ref{d1statsnorm} show that the distribution
of SNe~Ia ($\left\langle d_1 \right\rangle = 0.46 \pm 0.08$) has a
symmetrical appearance to the middle of spiral arms.
On the other hand, 
the distribution of SNe~Ibc ($\left\langle d_1 \right\rangle = 0.42 \pm 0.05$)
and SNe~II ($\left\langle d_1 \right\rangle = 0.43 \pm 0.03$)
are (respectively marginally and significantly) shifted towards the inner edges of arms.
The excess of SNe~Ibc toward inner edges of arms was also reported by
\citet{2007AstL...33..715M}. They pointed out that the distribution of SNe~Ibc
follows an exponential law relative to inner edges of spiral arms.
However, they fitted with
an exponential function the distribution of physical distances
of SNe from inner edges, and reported a scale length of 1~kpc.
The use of physical distances can significantly bias the results, because
spiral arms display a variety of sizes and shapes.

We now check the consistency of SN locations relative to the spiral
arms of their host galaxies, $d_1$, 
with the Gaussian function with mean $\mu = 0.5$. For this purpose
we perform a maximum likelihood estimation to find
the standard deviation of the Gaussian.
We then compare the distribution of $d_1$ distances
with the best fit function.
The corresponding KS and AD probabilities,  $P_{\mu=1/2}^{\rm KS}$ and $P_{\mu=1/2}^{\rm AD}$,
are given in Cols.~3 and 4 of Table~\ref{d1statsnorm}.

Then, we apply the same test, but without fixing the $\mu$
(the mean of the distributions).
The $P_{{\rm free}-\mu}^{\rm KS}$ and $P_{{\rm free}-\mu}^{\rm AD}$ values
(Cols.~5 and 6 in Table~\ref{d1statsnorm})
show that the distributions of all types of SNe
are consistent with a Gaussian function. However,
$P_{\mu=1/2}^{\rm AD} = 0.039$ value (Col.~4 in Table~\ref{d1statsnorm}) shows
that the distribution of SNe~II is not consistent
with a symmetric distribution around the middle of the spiral arm.
The inconsistency of
the distribution of SNe~Ibc with a symmetric one around the middle of the spiral
arm   is not significant ($P_{\mu=1/2}^{\rm KS} = 0.280$), probably because
of small number statistics.
The distribution of SNe~Ia is not shifted ($P_{\mu=1/2}^{\rm KS} = 0.967$)
with respect to the middle.

When comparing the distribution of $d_1$ distances of various SN types
(Cols. 2 and 3 in Table~\ref{pvalues}),
in contrast to the KS test ($P_{\rm KS}~=~0.150$),
the AD test shows significant difference
between the distributions of SNe~Ia and II ($P_{\rm AD}~=~0.013$).
The difference between the results of tests is due to the
difference of distributions in the tails (see also Fig.~\ref{distr1}).
The difference between the distributions of
SNe~Ia and Ibc is not significant,
probably because of the small number statistics.
The distributions of SNe~Ibc and II are not different
as well.

We analyze next the relative positions of different SN types
using the second normalization scheme presented in Sect.~\ref{normcor}.
These distributions are shown in the bottom panel of Fig.~\ref{distr1}.
The comparison of distributions of various types of SNe in the upper and bottom panels
of Fig.~\ref{distr1} shows that with the second normalization, the maximum
of SNe~Ibc distribution is shifted from inner edges of spiral arms
to their peaks (see also Table~\ref{d2statsnorm}).

As for the first normalization scheme,
we check the normality of the distribution
of $d_2$ distances.
The corresponding $P_{\mu=0}^{\rm KS}$, $P_{\mu=0}^{\rm AD}$,
$P_{{\rm free}-\mu}^{\rm KS}$ and $P_{{\rm free}-\mu}^{\rm AD}$ values
of KS and AD tests are given in Cols. 3, 4, 5 and 6
of the Table~\ref{d2statsnorm}.
These values show that, in contrast to $d_1$,
the distributions of $d_2$
for all types of SNe are consistent
with a Gaussian centred on the peaks of spiral arms.
Therefore, the asymmetric distributions of SNe~Ibc
\citep[obtained by][]{2007AstL...33..715M} and II with respect to the middle of spiral arm
are just an effect of first normalization scheme of SNe distances,
which does not take into account the intrinsic structure of spiral arms.

The differences between the distributions
presented in th bottom panel of Fig.~\ref{distr1}
again are checked by KS and AD tests (Cols. 4 and 5 in Table~\ref{pvalues}).
The tests show that there is a marginally significant difference
(probably, because of small number statistics)
between Ia and Ibc ($P_{\rm AD} = 0.054$)
as well as a significant difference between Ia and II
($P_{\rm AD} = 0.005$) SNe distributions.
The difference between the distributions of SNe~Ibc and II
is not statistically significant.

\begin{table}
  \centering
  \caption{Comparison of the distributions of the $d_1$ and $d_2$ distances
					 between different types of SNe.
					 The \textit{P}-values of both KS and AD tests are given.
					 The statistically significant differences are highlighted in bold.}
  \label{pvalues}
  \tabcolsep 5pt
  \begin{tabular}{lccccc}
    \hline
	  & \multicolumn{2}{c}{$d_1$} & & \multicolumn{2}{c}{$d_2$} \\
		\cline{2-3} \cline{5-6}
		Samples & $P_{\rm KS}$ & $P_{\rm AD}$ & & $P_{\rm KS}$ & $P_{\rm AD}$ \\
		\hline
    Ia versus Ibc & 0.109 & 0.073 & & 0.137 & 0.054 \\
		Ibc versus II & 0.835 & 0.780 & & 0.921 & 0.829 \\
		II versus Ia & 0.150 & $\textbf{0.013}$ & & 0.068 & \textbf{0.005} \\
		\hline \\
  \end{tabular}
\end{table}

Analyzing the distribution parameters of $d_2$
for various SN types presented in Table~\ref{d2statsnorm},
we note that the distributions of all types of SNe
show a significant concentration around the peak
of the spiral arm. The most concentrated is the distribution
of SNe~Ibc, then Types II and Ia.
In addition, the distributions of $d_2$ distances for all SN types are Gaussian,
without a significant shift from the peak of spiral arm profile.
The observed large numbers of SNe~Ia in interarm regions (Fig.~\ref{distr1})
is attributed to their long-tailed distribution.

Thus, we show that the intrinsic structure of spiral arms play a crucial role
shaping the distribution of CC~SNe relative to arms.
Therefore, in the remainder of this study we will present and discuss
only the distributions of $d_2$ distances.

\subsection{Distribution of SNe in GD and NGD hosts}
\label{grandfloc}

The influence of density waves on the star formation
in spiral galaxies has been one of the most debated problems,
with contradicting observational results \citep[for a recent review, see][]{2014PASA...31...35D}.
Due to the short lifetimes of their progenitors
\citep[e.g.][]{2006A&A...453...57J,2009ARA&A..47...63S,
2012MNRAS.424.1372A,2013MNRAS.428.1927C},
different types of CC SNe, considered as tracers of
star formation, can help to better understand
the influence of density waves on star formation.
Particularly, the comparative distribution of different types
of SNe relative to spiral arms of the hosts can shed light
on the nature of star formation in galaxies with different spiral structures. 

In our sample of 187 SNe host galaxies, 89 are GD
and 98 are NGD spirals. The distribution of SN types
in GD and NGD spirals is presented in Table~\ref{SNinAC}.
Because of the small number of SNe~Ibc in both samples
(10 and 15 SNe~Ibc in GD and NGD galaxies, respectively),
they are merged with SNe~II in the general class of CC SNe.

\begin{table}
  \centering
  \caption{Distribution of SN types according to
  the host spiral arm classes.}
  \label{SNinAC}
  \tabcolsep 5pt
  \begin{tabular}{lccr}
  \hline
  &\multicolumn{1}{c}{GD spirals}
  &\multicolumn{1}{c}{NGD spirals}&\multicolumn{1}{r}{All}\\
  \hline
    Ia & 29 & 36 & 65 \\
    Ibc & 10 & 15 & 25 \\
    II & 67 & 58 & 128 \\
  \hline
    All & 106 & 109 & 215 \\
  \hline \\
  \end{tabular}
\end{table}

\begin{figure}
\centering
  \includegraphics[width=\hsize]{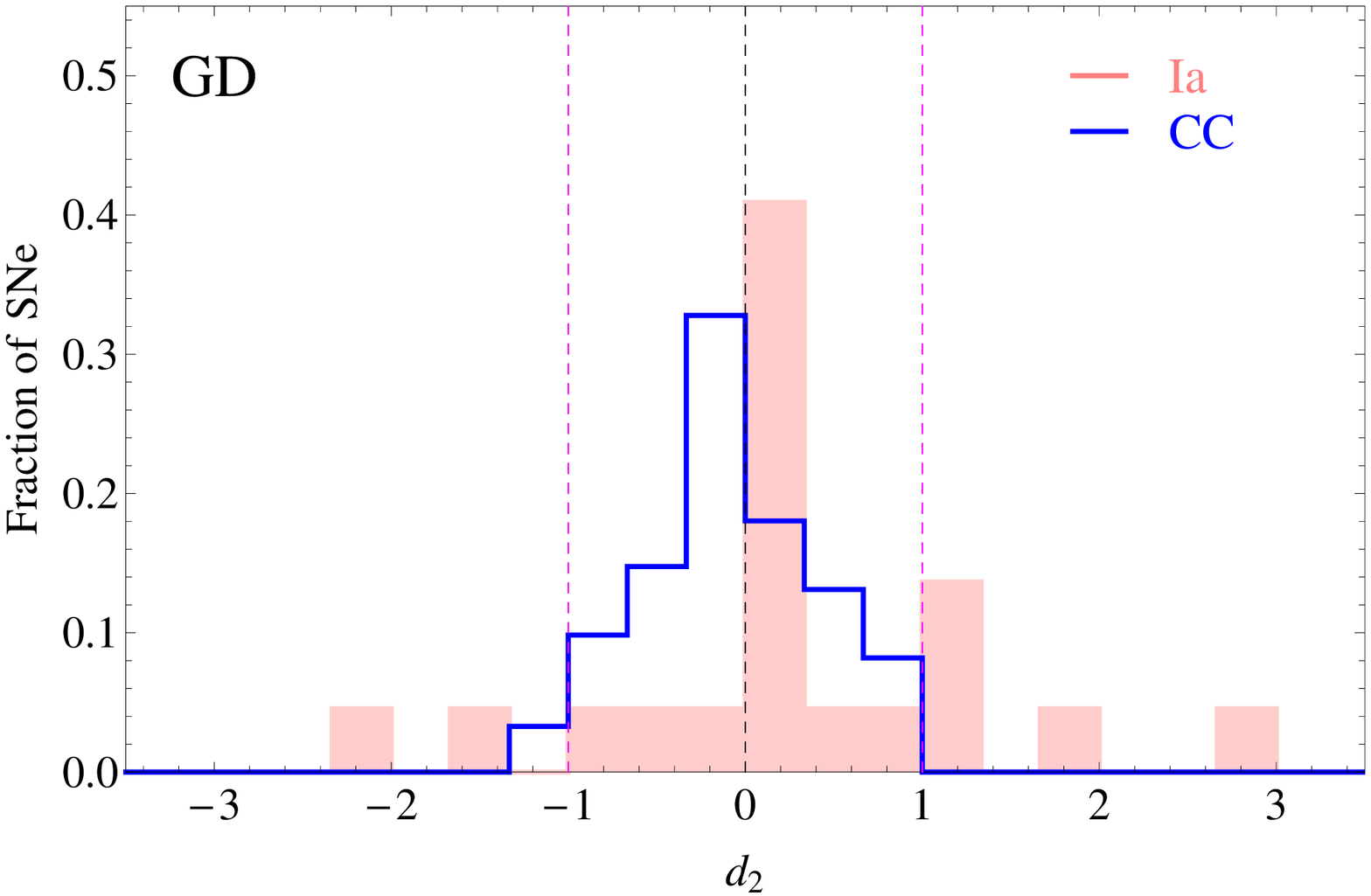}
  \includegraphics[width=\hsize]{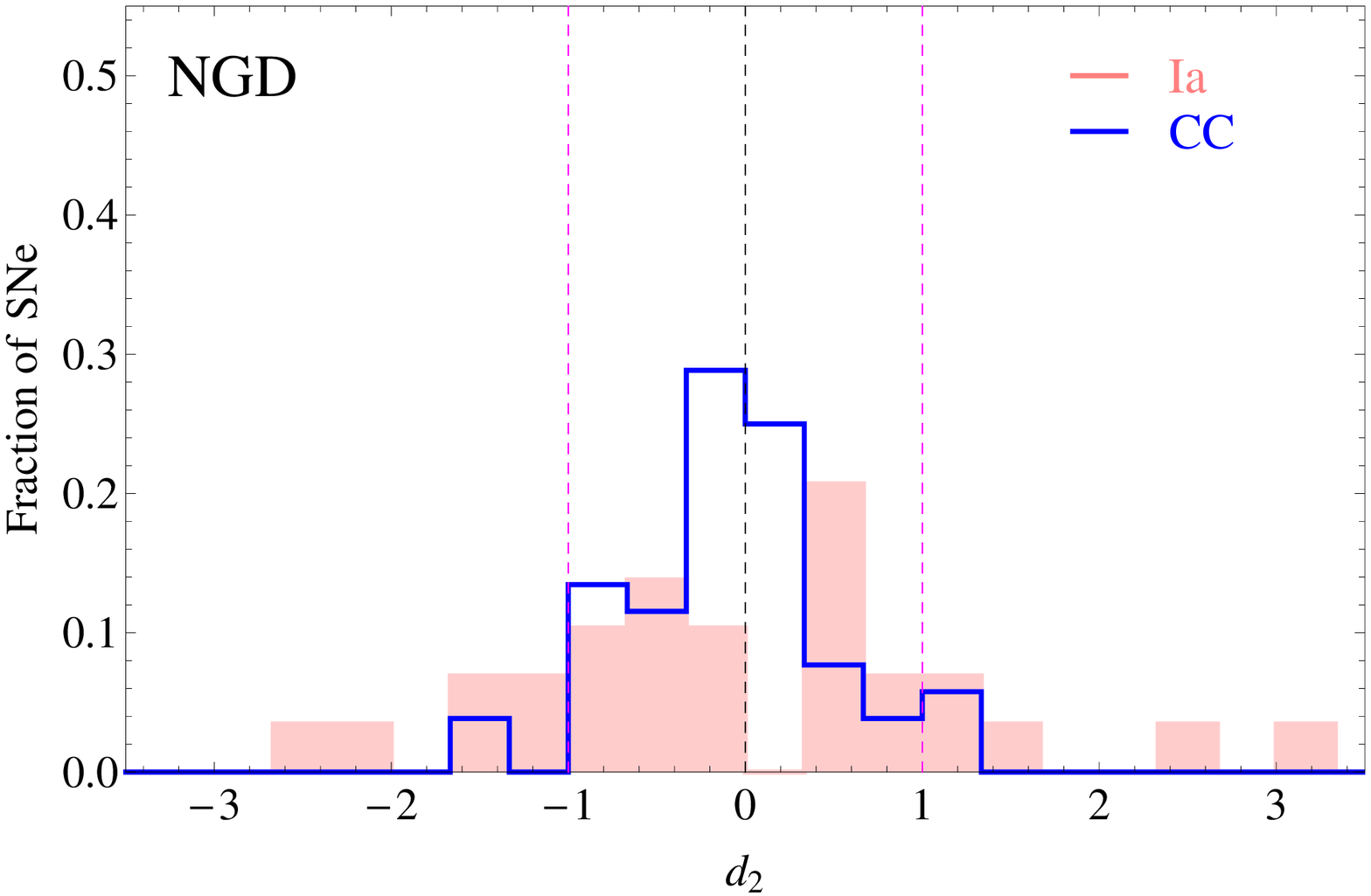}
  \caption{\emph{Upper panel}: 
distributions of the locations of 
SNe~Ia (filled histogram, 22 SNe) and CC (open histogram, 63 SNe) SNe
					relative to the peaks of spiral arms of GD galaxies. 
\emph{Bottom panel}: distributions of the locations of
					SNe~Ia (29 SNe) and CC~SNe (53 SNe) SNe relative to the
                    peaks of spiral arms of NGD galaxies. 
					SNe with distance flags `*' and `:' are removed.}
  \label{GDnGDplots}
\end{figure}

The distribution of Ia and CC SNe relative to the peaks of spiral arms
in GD and NGD galaxies are presented in Fig.~\ref{GDnGDplots}.
The mean and standard deviation,
with their bootstrapped ($10^{4}$ times) errors,
as well the $P_{\mu = 0}^{\rm KS}$ and
$P_{{\rm free}-\mu}^{\rm KS}$
values of KS normality tests are given in Table~\ref{dtildedp2}.
Here, the results of the AD test are not different from those of the KS test.
The $P_{\mu = 0}^{\rm KS}$ and $P_{{\rm free}-\mu}^{\rm KS}$ values show
that the distributions of CC and Ia SNe in both types of host galaxies
are consistent with Gaussian distributions centred
on the peaks of spiral arms (Table~\ref{dtildedp2}).

The results of the comparison of the distributions of $d_2$
of Ia and CC SNe in GD and NGD galaxies is given in Table~\ref{pvalac}.
The KS and AD tests show that, in both types of host galaxies,
the $d_2$ radial positions of Type Ia and CC SNe
within the spiral arms are significantly different.
The KS probability that such differences could arise by chance
are 0.012 (GD hosts) and 0.023 (NGD hosts).

\begin{table}
  \centering
  \caption{The mean, standard deviation of
	         $d_2$ distances, $P_{\rm KS}$ values
					 of normality tests.}
  \label{dtildedp2}
  \tabcolsep 5pt
  \begin{tabular}{lrrcc}
    \hline
    & \multicolumn{1}{c}{$\left\langle d_2 \right\rangle$} & \multicolumn{1}{c}{$\sigma$} &
		\multicolumn{1}{c}{$P_{\mu = 0}^{\rm KS}$} & \multicolumn{1}{c}{$P_{{\rm free}-\mu}^{\rm KS}$} \\
		\hline
   Ia$^{\rm GD}$ & $0.06 \pm 0.29$ & $1.37 \pm 0.29$ & 0.123 & 0.182 \\
	 Ia$^{\rm NGD}$ & $-0.03 \pm 0.24$ & $1.27 \pm 0.19$ & 0.972 & 0.972 \\
	 CC$^{\rm GD}$ & $-0.09 \pm 0.06$ & $0.51 \pm 0.04$ & 0.188 & 0.842 \\	
	 CC$^{\rm NGD}$ & $-0.10 \pm 0.08$ & $0.59 \pm 0.06$ & 0.169 & 0.558 \\
		\hline \\
  \end{tabular}
\end{table}

Analyzing both panels of Fig.~\ref{GDnGDplots}, we see that,
in both types of host galaxies, CC SNe are concentrated
towards the peaks of spiral arms. For SNe~Ia, a concentration
close to the peak of spiral arm in GD galaxies is observed,
while not in NGD galaxies, which actually show
a remarkable absence in the corresponding peak.
Finally, comparison of CC (as well as Ia) SNe distributions in GD and NGD galaxies
shows no significant difference (see Table~\ref{pvalac}).

\begin{table}
  \centering
  \caption{Comparison of the distributions of the $d_2$ distances
					 of various types of SNe in various arm classes of galaxies,
					 using the KS and AD tests.
					 The statistically significant differences are highlighted in bold.}
  \label{pvalac}
  \tabcolsep 5pt
  \begin{tabular}{lcc}
    \hline
	Samples & $P_{\rm KS}$ & $P_{\rm AD}$ \\
 		\hline
    Ia$^{\rm GD}$ versus CC$^{\rm GD}$ & \textbf{0.012} & \textbf{0.010} \\
		Ia$^{\rm NGD}$ versus CC$^{\rm NGD}$ & \textbf{0.023} & $\textbf{0.025}$ \\
		Ia$^{\rm GD}$ versus Ia$^{\rm NGD}$ & 0.231 & 0.461 \\
		CC$^{\rm GD}$ versus CC$^{\rm NGD}$ & 0.981 & 0.847 \\
		\hline \\
  \end{tabular}
\end{table}

It is interesting to note, that the distribution of CC SNe in NGD galaxies
is more extended ($\sigma = 0.59 \pm 0.06$)
than that in GD galaxies ($\sigma = 0.51 \pm 0.04$),
although the difference is not statistically significant.
There are more ($16^{+3}_{-3}$ per cent) CC~SNe (all of which are of Type~II)
in interarm regions of NGD than in interarm regions of GD galaxies ($9^{+3}_{-2}$ per cent).
From Table~\ref{dtildedp2}, we note that the sigmas ($\sigma$) of the
distributions of SNe~Ia in GD and NGD galaxies are consistent. The
differences between the distributions of SNe~Ia and CC~SNe in both GD and
NGD galaxies are attributed to the remarkably extended distributions of
SNe~Ia.

\subsection{Radial behaviour of SNe distributions in spiral arms}
\label{radtrend}

The spatial distribution of SNe in host galaxies provides
strong constraints on the nature of SN progenitors as well
establishes the relations with stellar population of the galaxies.
Various studies show that CC SNe are tightly connected
to discs \citepalias[e.g.][]{2016MNRAS.456.2848H} and spiral arms \citep[e.g.][]{1996AJ....111.2017V},
which differs from the distribution of SNe~Ia in the arms \citep[e.g.][]{1996ApJ...473..707M}.
In addition, studying the radial distributions of the sample of SNe~Ibc and II,
several authors have found that the mean galactocentric distance of SNe~Ibc is
smaller than that of SNe~II (e.g. \citealt{1997AJ....113..197V,
2009A&A...503..137B,2009MNRAS.399..559A,2009A&A...508.1259H}; \citetalias{2016MNRAS.456.2848H}).

\begin{figure}
\centering
  \includegraphics[width=\hsize]{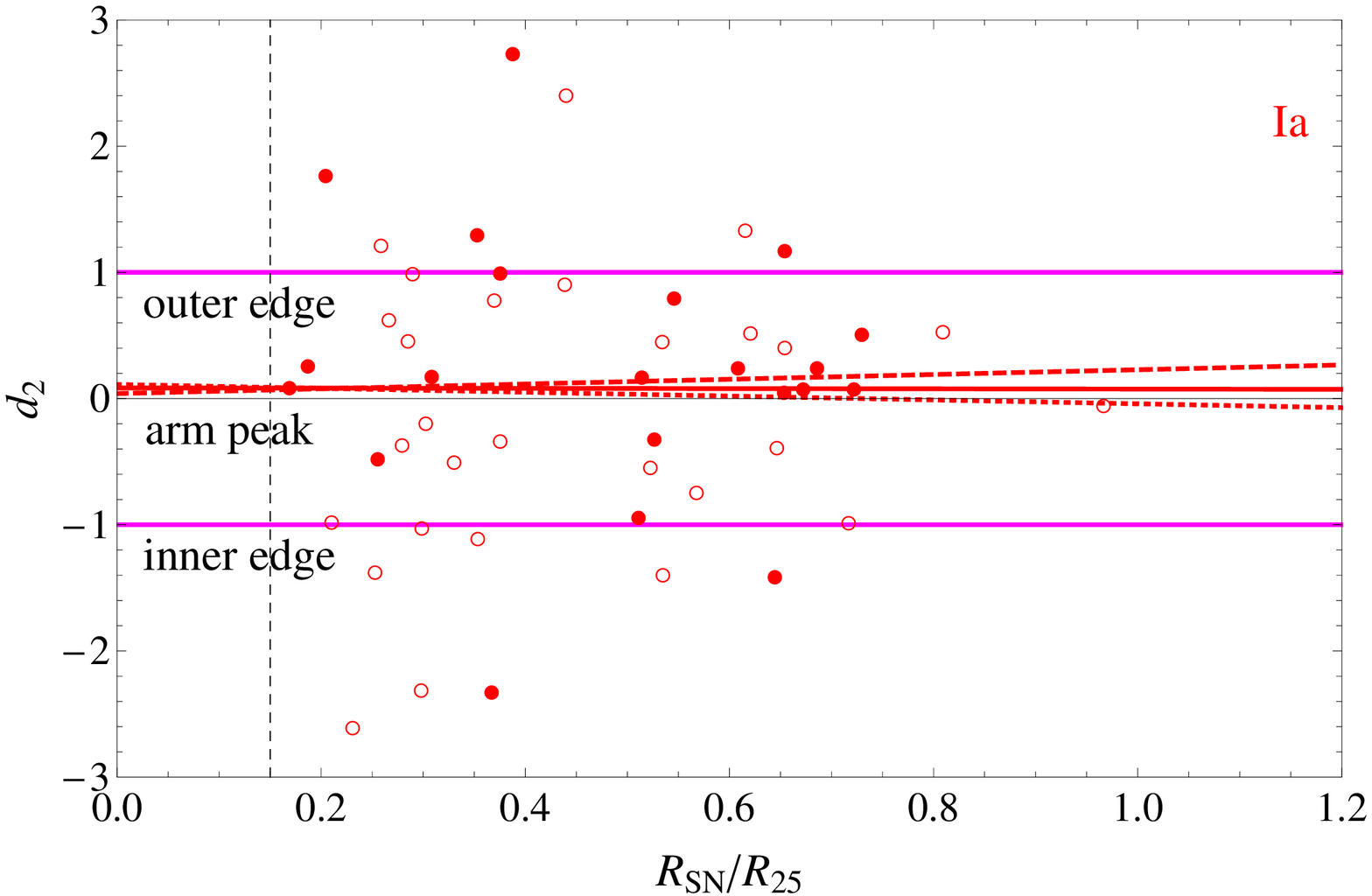} \\
  \includegraphics[width=\hsize]{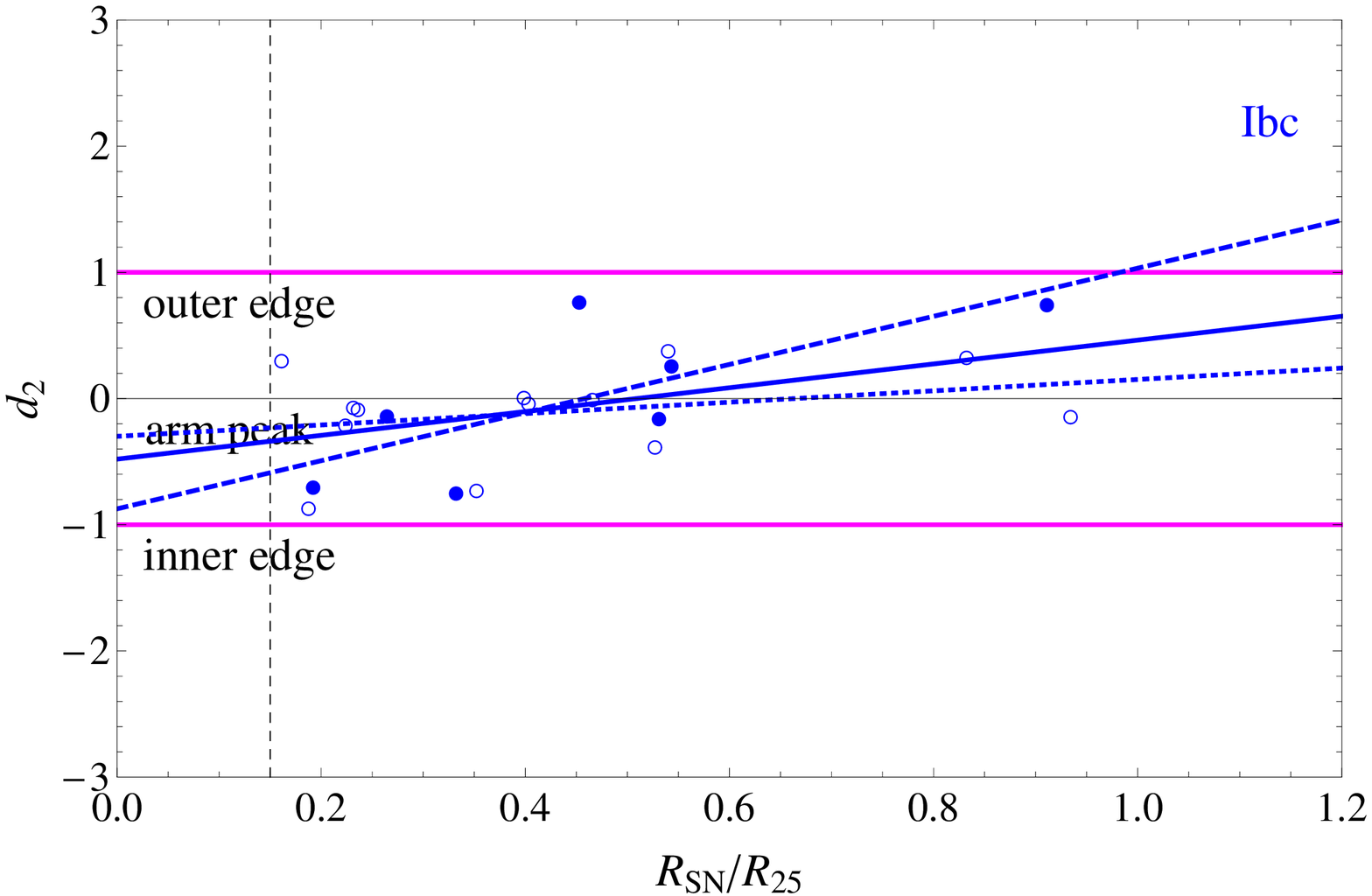} \\
  \includegraphics[width=\hsize]{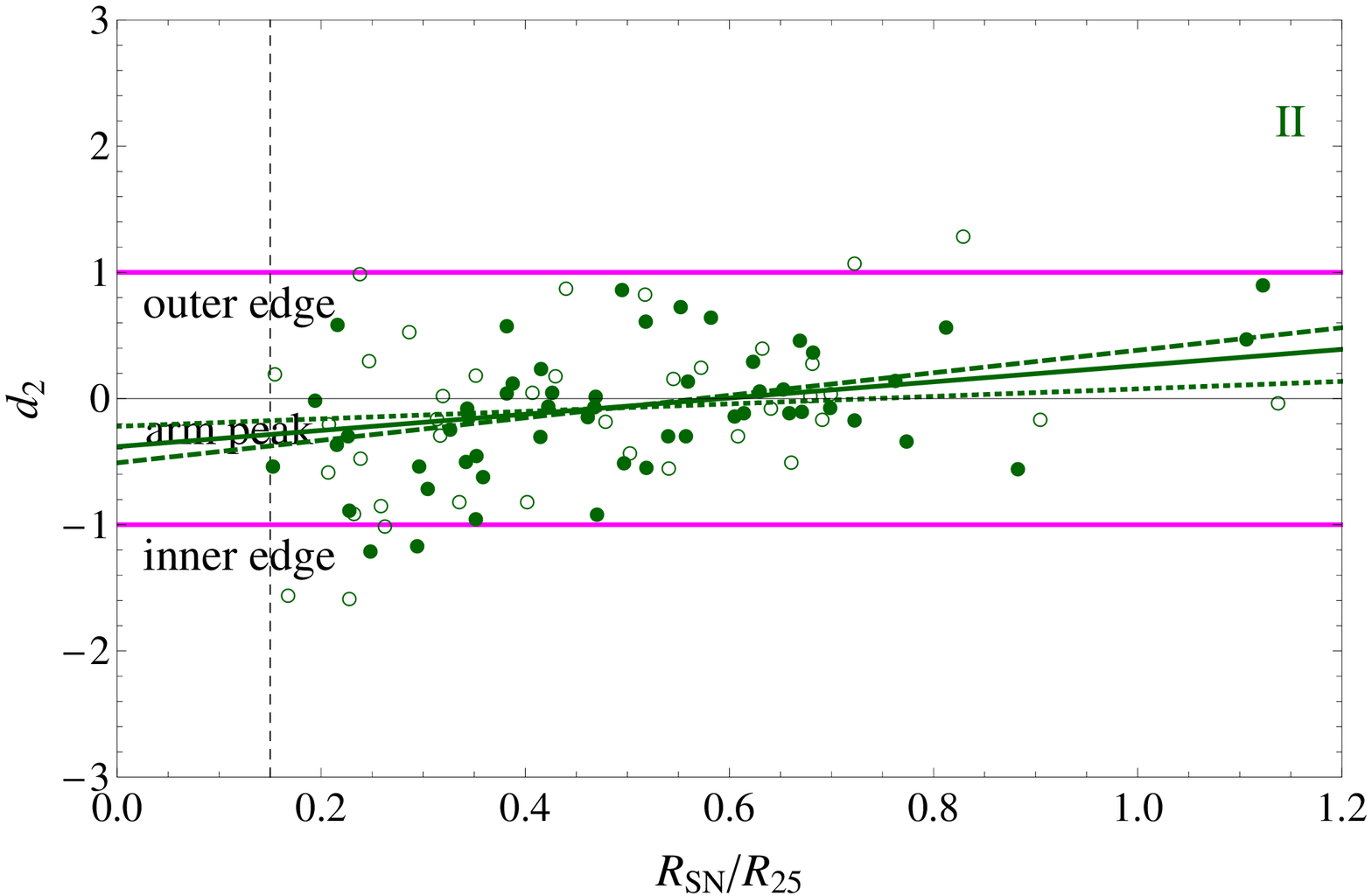} \\
  \caption{Distributions of the $d_2$ distances of SNe~Ia (upper),
	         Ibc (middle), and II (bottom) relative to the peaks of
		       spiral arms versus the deprojected and normalized galactocentric distance.
					 The data flagged by the ``*'' and ``:'' symbols are not considered.
					 Filled and open circles respectively  show SNe in GD and NGD galaxies.
		       In all figures, the linear fits are given
					 for the full (solid), GD (dashed), and NGD (dotted) samples.
		       The inner and outer edges, as well the peaks
					 of spiral arms are shown with parallel lines,
		       similar to Fig.~\ref{profile}. The region to radius $R_{\rm SN}/R_{\rm 25} = 0.15$
			     (dashed vertical line) have been excluded from this analysis.}
  \label{peakbrightdist}
\end{figure}

Therefore, we investigate the possible dependence
of $d_2$ on the deprojected radii
($R_{\rm SN}$) of the SNe, normalized to the $R_{\rm 25}$ of their hosts
($R_{\rm SN}/R_{\rm 25}$). Fig.~\ref{peakbrightdist} presents the
radial trend for various SN types.
Since we aim to investigate the possible influence of density waves
on the distribution of SNe in spiral arms,
here we consider only \textit{arm} SNe,
although, for better visualization, the interarm SNe
are also shown in Fig.~\ref{peakbrightdist}.
In the full (GD+NGD) sample of hosts for both types
of CC SNe, there is a positive correlation between $d_2$
and $R_{\rm SN}/R_{\rm 25}$ (see solid trendlines
in the middle and bottom panels of Fig.~\ref{peakbrightdist}).
Both SNe~Ibc and II at smaller radii are enhanced in the inner parts of spiral arms
(before the peak of spiral arm), and at larger radii the enhancement
is shifted to the outer parts (after the peak of spiral arm).

Interestingly, Figure~\ref{peakbrightdist} indicates that 
the trendlines of SNe~Ibc and II cross the
arm peak at approximately the same galactocentric distances.
Noting that the corotation radii reported in \citet{2013MNRAS.428..625S},
normalized 
to the $R_{\rm 25}$ of each galaxy, obey
$\left\langle r_{\rm cor}/R_{\rm 25}\right\rangle~\approx~0.45~\pm~0.03$, we
find that the trendlines of SNe~Ibc and II cross the
arm peak at roughly the location of
the corotation radius.

\begin{table*}
  \centering
  \begin{minipage}{167mm}
  \caption{Spearman's rank correlation $P$-values
	         of the radial trend of SNe distribution
					 relative to the peaks of spiral arms.
					 The statistically significant trends
					 are highlighted in bold.}
  \label{rdpspear}
  \tabcolsep 5pt	
	\begin{tabular}{lcrccccrccccrcc}
    \hline
		 SNe & \multicolumn{4}{c}{All spirals} &  & \multicolumn{4}{c}{GD galaxies} &  & \multicolumn{4}{c}{NGD galaxies} \\
		\cline{2-5} \cline{7-10} \cline{12-15}
    & $N_{\rm SN}$ & \multicolumn{1}{c}{a} & \multicolumn{1}{c}{$\rho$} & \multicolumn{1}{c}{P} &  &
		$N_{\rm SN}$ & \multicolumn{1}{c}{a} & \multicolumn{1}{c}{$\rho$} & \multicolumn{1}{c}{P} &  &
	  $N_{\rm SN}$ & \multicolumn{1}{c}{a} & \multicolumn{1}{c}{$\rho$} & \multicolumn{1}{c}{P} \\
		\multicolumn{1}{c}{(1)} & \multicolumn{1}{c}{(2)} & \multicolumn{1}{c}{(3)} &
		\multicolumn{1}{c}{(4)} & \multicolumn{1}{c}{(5)} &  & \multicolumn{1}{c}{(6)} &
		\multicolumn{1}{c}{(7)} & \multicolumn{1}{c}{(8)} & \multicolumn{1}{c}{(9)} &  &
	  \multicolumn{1}{c}{(10)} & \multicolumn{1}{c}{(11)} & \multicolumn{1}{c}{(12)} & \multicolumn{1}{c}{(13)} \\
		\hline
    Ia & 34 & $-0.01 \pm 0.72$ & 0.013 & 0.942 &  & 15 & $0.19 \pm 0.58$ & 0.086 & 0.761 &  &
		19 & $-0.15 \pm 0.82$ & -0.096 & 0.694 \\
		Ibc & 20 & $0.94 \pm 0.39$ & 0.465 & \textbf{0.039} &  & 7 & $1.91 \pm 0.67$ & 0.571 & 0.180 &  &
		13 & $0.45 \pm 0.41$ & 0.302 & 0.316 \\
		II & 87 & $0.64 \pm 0.22$ & 0.291 & \textbf{0.006} &  & 52 & $0.89 \pm 0.27$ & 0.385 &
		\textbf{0.005} &  & 35 & $0.30 \pm 0.38$ & 0.168 & 0.336 \\
		\hline \\
  \end{tabular}
	\end{minipage}
\end{table*}

To check the significance of the radial trends, we use the
Spearman's rank correlation test. For the full sample of the hosts,
the \textit{P}-values of the hypothesis that there is no
dependence between $d_2$ and $R_{\rm SN}/R_{\rm 25}$
are presented in Column~5 of Table~\ref{rdpspear}.
The number of SNe, \textit{a} coefficient
of the $d_2 = a\,(R_{\rm SN}/R_{\rm 25}) + b$ linear interpolation
between $d_2$ and $R_{\rm SN}/R_{\rm 25}$, and $\rho$ parameter
of Spearman's rank correlation test are given
in Cols.~2, 3 and 4 of Table~\ref{rdpspear}, respectively.
Table~\ref{rdpspear} shows
that the trends for both SNe~Ibc and II
in the full sample are significant.

To strengthen the results, we merge
together SNe~Ibc and SNe~II into a CC~SNe sample.
We use the $\left\langle r_{\rm cor}/R_{\rm 25}\right\rangle$
demarcation radius to separate the samples of SNe into two radial bins.
The KS and AD tests show that the difference of $d_2$ distributions
of CC SNe in these radial bins is statistically significant
($P_{\rm KS}~=~0.036$ and $P_{\rm AD}~=~0.010$).

Summarizing the results for SNe~Ibc and II in  the full sample,
we conclude that: (1) there are significant positive correlations between
$d_2$ and $R_{\rm SN}/R_{\rm 25}$;
(2) in most cases, the positions of CC SNe are
inside (outside) the peak of the spiral arms
for galactocentric radii smaller (larger) than the mean corotation radius,
$\left\langle r_{\rm cor}/R_{\rm 25}\right\rangle~\approx~0.45~\pm~0.03$.
We will discuss these results below merging them
with appropriate results obtained for GD and NGD subsamples.

We now investigate the existence of the possible radial trends of $d_2$ distributions
for SNe~Ibc and II  separately in GD and NGD galaxies.
The Spearman's rank correlation \textit{P}-values
of the radial trends are presented in Cols.~9 and 13 of Table~\ref{rdpspear}.
The definition of Cols. 6, 7, 8 and 10, 11, 12 is similar to Cols. 2, 3 and 4,
but represent the values in GD and NGD galaxies, respectively.
Here, the small number statistics for SNe~Ibc probably play a role.
Having this in mind, a positive significant correlation
is observed  in GD hosts for SNe~II
between $d_2$ and $R_{\rm SN}/R_{\rm 25}$. An even stronger  correlation
($\rho=0.57$) occurs
for SNe~Ibc, but it is not statistically significant probably due to the small number
statistics (only 7 SNe~Ibc in GD hosts). In contrast, in NGD galaxies we find
no significant correlations between 
$d_2$ and $R_{\rm SN}/R_{\rm 25}$ 
for SNe~Ibc and II.

Similarly to the full sample of host galaxies, we split the CC SNe sample
into two radial bins
using the demarcation radius of $\left\langle r_{\rm cor}/R_{\rm 25} \right\rangle$,
for SNe in both GD and NGD host galaxies, and compare the distributions
of their $d_2$ parameters. The results
of KS and AD tests for GD and NGD hosts
are $P_{\rm KS} = 0.082$ ($P_{\rm AD} = 0.024$) and
$P_{\rm KS} = 0.496$ ($P_{\rm AD} = 0.281$), respectively.
This means that, in contrast to NGD galaxies, the density waves
of GD galaxies can strongly act on the star formation processes,
thus exhibiting significant differences between distributions of
CC~SNe relative to peaks of spiral arms inside and outside the corotation radius.
Most probably, the absence of such a distribution in the younger stellar
component of NGD galaxies is due to the weaker nature
of the density waves in NGD spirals.

In contrast to CC SNe, Fig.~\ref{peakbrightdist} and Table~\ref{rdpspear}
show that the distribution of SNe~Ia in spiral arms does not
depend on the galactocentric radius in the full sample,
as well as in both GD and NGD galaxies separately.
Similar to CC SNe, we split the SNe~Ia sample into two radial bins
with the demarcation radius and compare the $d_2$ distributions
of these SNe inside inner and outer radial bins. The differences 
between the two distributions in the
full ($P_{\rm KS} = 0.882$ and $P_{\rm KS} = 0.720$),
GD ($P_{\rm KS} = 0.919$ and $P_{\rm AD} = 0.722$)
and NGD ($P_{\rm KS} = 0.904$ and $P_{\rm AD} = 0.815$) samples are not significant.
It is important to note that the number of interarm SNe~Ia
is enhanced at smaller radii (see Fig.~\ref{peakbrightdist}),
which is probably due to the
contribution of the old bulge component.
However, we do not exclude that a significant part of these
interarm SNe~Ia belong to the disc population \citepalias[][]{2016MNRAS.456.2848H}.

\section{Discussion}
\label{discus}

The results  described above deserve a deeper analysis in order to fully
understand all the possible implications in the context of the host
galaxy spiral structure and SN progenitor scenario.

Although the difference between the distributions of $d_2$ distances
of the locations of SNe relative to the peak of spiral arms,
normalized to the inner/outer semi-widths)
of CC~SNe in GD and NGD galaxies is not statistically significant (Sect.~\ref{grandfloc}),
the radial dependence of these distributions in GD and NGD galaxies
have different behaviors (Sect.~\ref{radtrend}). In particular, in GD galaxies,
a positive significant correlation exists between
the galactocentric radius of SNe~II and $d_2$, and an even stronger (but not significant)
correlation occurs for SNe~Ibc.
In contrast, in NGD galaxies there are no significant correlations between these parameters.
Since the progenitors of SNe~Ibc and II belong to young stellar populations,
we interpret the differences as the consequence of star formation processes
in spiral arms and their differences between GD and NGD galaxies.

In NGD galaxies, the spiral arms are believed
to originate from the shear of HII regions due to the differential rotation of the discs
(e.g. \citealt{1982FCPh....7..241S,2003ApJ...590..271E}; \citealt*{2011MNRAS.417.1318D}).
In contrast to GD galaxies, their spiral arms corotate
with the discs and do not show signs of shocks
in their leading edges \citep[see review by][]{2014PASA...31...35D}.
Therefore, one does not expect young stars to be
concentrated towards one of the edges of arms in NGD galaxies.
This scenario also predicts
the absence of radial trends for the distributions of SNe~Ibc and II
inside the spiral arms. In NGD galaxies, we found that
CC SNe are more concentrated towards the peaks
of spiral arms than SNe~Ia.
Moreover, despite small number statistics, in NGD galaxies
we found different concentration levels for SNe~Ibc and II.
In particular the mean absolute $d_2$ distance
of SNe~Ibc is $0.27 \pm 0.08$ ($N = 13$)
and for SNe~II is $0.49 \pm 0.07$ ($N = 39$).
An AD test shows that the difference between the distributions
of absolute distances of SNe~Ibc and II from the peaks of spiral arms
is barely significant ($P_{\rm AD} = 0.074$, while $P_{\rm KS} = 0.207$).
Hence, in NGD galaxies the shortest mean distance
to the peak of spiral arm is for SNe~Ibc.
In addition, the distribution of any SN type inside the spiral arms
in NGD galaxies does not show any significant radial trend.

Assuming that 1) the \textit{g}-band profiles of spiral arms
represent the distribution of young stars, and 2) the
peaks of spiral arms of NGD galaxies are the most suitable sites of the
star formation, we propose that when the concentration
of a given type of SNe towards the arm peak is higher,
their progenitors are younger (more massive in the context of single-star evolution).
Thus, a mass sequence Ia--II--Ibc for the SN progenitors is expected,
in agreement with those from the literature obtained by
the association of various SN types with the
$\rm H\alpha$ emission of the host galaxy
\citep[e.g.][]{2006A&A...453...57J,2008MNRAS.390.1527A,2012MNRAS.424.1372A}.
This also confirms the result of \citet*{2008ApJ...687.1201K}
for the local environment of the SNe,
that SNe~Ibc prefer to lie in relatively bright regions, in comparison with SNe~II and Ia.

The distribution of SNe inside the spiral arms
of GD galaxies is quiet different.
In particular, in GD galaxies we found a significant shift between
the distributions of distances of CC and Ia SNe from the peaks of spiral arms.
In order to understand the possible effect of density waves on the
observed offset, we pay attention to the fig.~4 of \citet{1998A&A...340....1D},
representing the effect of a strong density wave on the \textit{B}-
and \textit{I}-band profiles of spiral arms before corotation radius.
This figure shows that the strong density waves
that cause the young stars to concentrate
in the inner (leading) edges of spiral arms,
lead to an intensity (brightness) profile that is skewed
to the inner regions of spiral arms
in the short wavebands ($B$ and $g$)
in comparison to the longer ones ($I$, e.g. \citealt{1998A&A...340....1D}; \citealt*{2009ApJ...694..512M}).
This effect is schematically illustrated in fig.~1 of \citet{2009ApJ...694..512M}.
It is probably also reflected on the corresponding distributions
of CC~SNe versus SNe~Ia in the upper panel of Fig.~\ref{GDnGDplots}.

\begin{figure}
  \begin{center}$
  \begin{array}{c}
  \includegraphics[width=0.9\hsize]{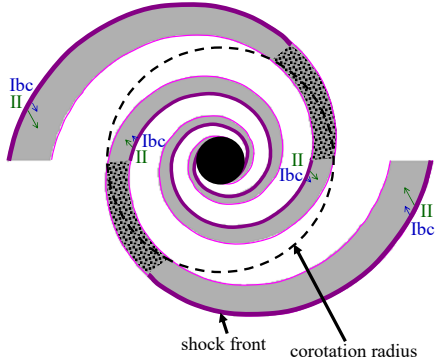}
  \end{array}$
  \end{center}
  \caption{The scheme of star formation distribution in a model of two armed GD galaxy
	         with the directions and relative sizes of drifts
           from birth places up to the explosion
					 for SNe~Ibc (blue arrow) and II (green arrow).
					 For better visualization, the directions of drifts are shown with
					 a significant radial component.}
  \label{GDchart}
\end{figure}
\begin{figure}
  \begin{center}$
  \begin{array}{c}
  \includegraphics[width=0.9\hsize]{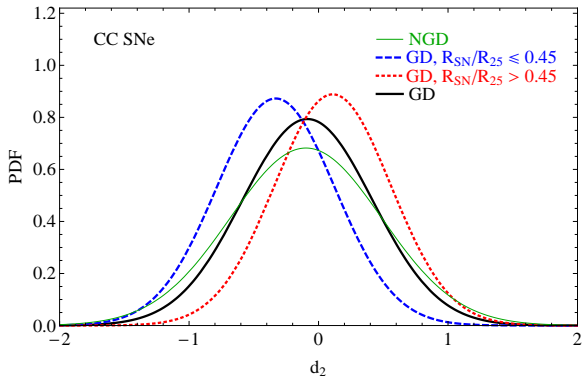}
  \end{array}$
  \end{center}
  \caption{The fitted Gaussian functions of CC SNe distributions in NGD (thin green), GD (thick black),
           as well in inner ($R_{\rm SN}/R_{\rm 25}~\leq~0.45$, dashed blue)
					 and outer ($R_{\rm SN}/R_{\rm 25}~>~0.45$, dotted red) galactocentric distances of GD galaxies.}
  \label{CCdistrfunc}
\end{figure}

In addition, we found an offset (though non-significant)
between the distributions of distances of SNe~Ibc and II from
the inner edges of the spiral arms. The observed offset between all types of SNe
is ordered with the following sequence
from inner edges Ibc-II-Ia (as in NGD galaxies). It is important to note
that in GD hosts, there is a statistically significant
positive correlation for \textit{arm} SNe~II
between $d_2$ and $R_{\rm SN}/R_{\rm 25}$.
The same correlation is found for SNe~Ibc,
with an even higher slope, but not significant
because of small number statistics.
The positions of both types of SNe beyond $R_{\rm SN}/R_{\rm 25}~\approx~0.45$
(roughly the mean corotation radius) are now typically outside the peaks of the arms
through the radius vector, while at smaller radii
the positions of SNe are typically inside the peaks of the arms.
Similar trends for star-forming regions are observed
in some GD galaxies \citep[e.g.][]{2013A&A...560A..59C}.

Adopting an average corotation radius of $0.45~R_{\rm 25}$,
the mean distances ($d_1$ inside and $1-d_1$ outside
the corotation radius, respectively)
of SNe~Ibc and II from leading edges of spiral arms
are $0.25 \pm 0.07$ and $0.44 \pm 0.03$, respectively. 
KS and AD tests show that the difference
between the distributions of SNe~Ibc and II
relative to leading edges of spiral arms
is statistically significant ($P_{\rm KS} = 0.026$ and $P_{\rm AD} = 0.011$).
According to the dynamical simulations by \citet{2010MNRAS.409..396D},
the observed concentration sequence towards the leading edges of spiral arms indicates
a lifetime sequence \citep[see the top-left panel of fig.~4 in][]{2010MNRAS.409..396D}
for their progenitors (from youngest to oldest).
Therefore, one may gain information about the lifetimes of SN
progenitors from the amplitude of the shift and the adoption
of an average corotation radius
(the greater the mean distance from the leading edge,
the longer is the progenitor's lifetime).
It is important to note that the real migration of stars from their birthplace
in the spiral arms is quite complex  (e.g. \citealt*{2013MNRAS.436.1479K}; \citealt{2015A&A...578A..58H})
and a detailed analysis of it is beyond the aim of this paper
that tackles the issue just qualitatively.

The scheme of star formation in a model of a GD galaxy with two spiral arms
with the directions and relative sizes of drifts
from birth places up to the explosion
for various SNe is given in the Fig.~\ref{GDchart}.
Inside the corotation radius (dashed circle), star
formation processes generally occur in a shock front
at the inner (leading) edges of spiral arms.
Since  the disc rotates faster than the spiral arms inside the corotation radius,
newborn stars near inner edges move towards
the outer edges of spiral arms. On the contrary, outside the corotation
radius, stars are caught up by the spiral arms, hence move from the outer
edges of the arms towards the inner edges of spiral arms.
In the corotation zone, there are no triggering mechanisms of star formation, such as
spiral shocks. The main mechanism of star formation in this region (dotted surface of arms) is
gravitation instability (as in NGD galaxies). Therefore, in this region, the distribution
of SNe inside the spiral arms should have the same behavior as in NGD galaxies.
Moreover, because of the absence of spiral shocks in this region,
star formation (e.g. \citealt*{1992ApJS...79...37E}),
hence the number of CC SNe, should exhibit a drop.
Since more massive stars live shorter than less massive ones,
their explosion sites are, on average, closer to the leading edges of arms
where they born.
The observed significantly shorter distances
of SNe~Ibc from the leading edges of spiral arms show
that their progenitors are younger (more massive) than those of SNe~II.
This result is in agreement with the single-star progenitor scenario of SNe~Ibc.

Since we cannot directly measure the corotation radii,
using instead isophotal radii to scale our host galaxies,
any dispersion of the corotation-to-isophotal radius ratio
will blur the expected radial drop near corotation
(see the continuous distribution in CC SN positions
within arms versus galactocentric radius in Fig.~\ref{peakbrightdist}).
Nevertheless, the observed intersection of the linear fits
to the normalized locations of CC SNe within arms
(in units of isophotal radii) with the peaks of arms
at precisely the typical corotation radii (again in units of isophotal radii)
supports our conclusion that the locations of SNe are a combination
of the circular velocity of stars in the disc relative
to the pattern speed of the spiral arms and the ages
of the progenitors.\footnote{The distribution of SNe
relative to the estimated corotation radii will be investigated
in our forthcoming paper, with about twice larger sample and
without restrictions on the environments of SNe in galaxies.}

Our results show that to constrain the nature of CC SN progenitors,
it is very important to take into account the spiral arm classes of host galaxies, as well
the locations of SNe relative to the spiral arms of their host galaxies. 
We illustrate this in Fig.~\ref{CCdistrfunc}, where we show the 
best-fit Gaussians to the CC~SNe distributions in NGD and
GD galaxies, as well the distributions 
in inner ($R_{\rm SN}/R_{\rm 25}~\leq~0.45$) and
outer ($R_{\rm SN}/R_{\rm 25}~>~0.45$) galactocentric distances.
Fig.~\ref{CCdistrfunc} shows that
the distribution of distances of CC SNe relative to the peaks
of spiral arms in NGD and GD (without radial separation) galaxies are
similar, repeating our finding of
Sect.~\ref{grandfloc} that the differences
of these distributions are not statistically significant.
However, when we separate the distribution in GD galaxies into two radial bins, they exhibit
an obvious offset between each other, as well from the peaks of spiral arms,
as we previously found in Sect.~\ref{radtrend}.
The difference of the distributions of CC SNe in inner and outer radial bins
of NGD galaxies is not statistically significant (Sect.~\ref{radtrend}).
This shows that the distributions of $d_1$ and $d_2$ distances of
SNe~Ibc and II in the whole sample of host galaxies
are the combinations of 3 different distributions:
the one in NGD galaxies;
the distribution at smaller ($R_{\rm SN}/R_{\rm 25}~\leq~0.45$) and
larger ($R_{\rm SN}/R_{\rm 25}~>~0.45$) galactocentric distances of GD galaxies.
Therefore, their combined distributions are less informative
for the ages of SN progenitors.
It is important to note that the previous studies
\citep[e.g.][]{1994PASP..106.1276B,2007AstL...33..715M}
suffer from this effect, and the obtained mean distances
of SNe from the arm peaks in joint samples
of host galaxies (GD+NGD) do not represent
the mean distances from the birth places of SNe.

We stress that the $d_2$ distribution of SNe~Ia relative to spiral arms of GD and NGD galaxies
exhibits a similar behavior. In both types of galaxies, these distributions are
significantly different from the distributions of CC SNe.
In addition, in both samples of galaxies these
distributions do not show any significant dependence on galactocentric distances.
This means that the progenitors of SNe~Ia arise from a population
that is old enough to migrate away (in both radial and azimuthal directions)
from their birth places during their lifetimes, thus
representing only the distribution of the older stellar population.

\section{Conclusions}
\label{concl}

In this fourth paper of our series of articles on the relation between SNe
and their host galaxies, we analyse the
distributions of different types of SNe relative to spiral arms
of galaxies with different arm classes. 
We use a large and homogeneous sample to study the nature
of SN progenitors, through the star formation processes
in spiral galaxies and their possible triggering mechanisms.
The sample of this analysis consists of 187 spiral
galaxies, which host 215 SNe in total.
Out of these 215 SNe, 106 occurred in GD galaxies and 109 in NGD galaxies.

The main results of the article and their interpretations
are summarized below.

\begin{enumerate}
  \item The most significant parameter for a quantitative
        analysis of the spatial distribution of SNe inside the
        spiral arms is $d_2$ (the distance of an SN
				from the peak of its spiral arm, normalized to the inner/outer semi-width),
				as defined in Sect.~\ref{Armposition} (eq.~[\ref{d2}]).
	\item For all SNe, the distribution of $d_2$ values is in agreement
				with a Gaussian, centred on zero (corresponding to the peak of the  arm).
				However, SNe~Ibc are most concentrated around the arm peaks,
        SNe~II less so, while SNe~Ia display the weakest
        concentration around the peaks.
	\item The distributions of SNe~Ia and CC~SNe relative to peaks of spiral arms
				are	significantly different from each other
				in both GD and NGD galaxies. Moreover, the peaks of
				the distributions of Type~Ia and CC~SNe in GD galaxies show offsets similar to
				the color gradients in spiral arms, as predicted from density wave theory
				\citep[e.g.][]{1998A&A...340....1D,2009ApJ...694..512M}.
	\item In NGD hosts, no significant correlation is observed
				between $d_2$ and $R_{\rm SN}/R_{\rm 25}$ for \textit{arm} SNe~Ibc and II.
				This is probably due to the nature of spiral arms of NGD galaxies,
				that reflects the distribution of massive star formation in arms of these hosts.
	\item In GD hosts, a statistically significant positive correlation is observed
				between $d_2$ and $R_{\rm SN}/R_{\rm 25}$ for \textit{arm} SNe~II.
				An even stronger correlation is present for SNe~Ibc, but is not statistically
				significant, due to their small number statistics.
				Most probably, because of the shorter lifetimes of their progenitors,
				the slope of the radial trend of SNe~Ibc
				is higher than that for SNe~II.
	\item In GD galaxies, the distribution
				of $d_2$ distances of CC SNe in two different radial bins
				(with demarcation radius $r_{\rm cor}/R_{\rm 25}~\approx~0.45$)
				are significantly different, while in NGD the difference is not significant.
	\item A similar comparison of the distributions of $d_2$
				of SNe~Ia with the same demarcation radius ($r_{\rm cor}/R_{\rm 25}~\approx~0.45$)
				in both GD and NGD galaxies shows no significant differences.
				Moreover, for SNe~Ia, $d_2$ shows
				no significant dependence on $R_{\rm SN}/R_{\rm 25}$.
	\item In GD galaxies, the distributions of distances
				of SNe~Ibc and II from the leading edges are significantly different.
				SNe~Ibc occur at smaller distances from the leading edges of arms
				than that of SNe~II.
				Hence, the $d_2$ versus $R_{\rm SN}/R_{\rm 25}$ correlation
				for SNe~Ibc and II in GD galaxies is in agreement with the density wave-induced
				star formation model and with the following lifetime sequence of
				SN progenitors: Ibc-II-Ia.
\end{enumerate}

The results of this study show that the distribution
of SNe relative to spiral arms is a powerful tool
to constrain the lifetimes (masses) of their progenitors
and to better understand the star formation processes
in various types of spiral galaxies.

\section*{Acknowledgments}

We would like to thank the referee, Phil~James, for
constructive comments that improved the clarity of this paper.
LSA, AAH, and ARP acknowledge the hospitality of the
Institut d'Astrophysique de Paris (France) during their
stay as visiting scientists supported by the Collaborative
Bilateral Research Project of the State Committee of Science (SCS)
of the Republic of Armenia and the French
Centre National de la Recherch\'{e} Scientifique (CNRS).
This work was supported by State Committee Science MES RA,
in frame of the research project number SCS~13--1C013.
AAH is also partially supported by the ICTP.
VA is supported by grant SFRH/BPD/70574/2010 from FCT (Portugal).
DK acknowledges financial support from the Centre
National d'\'{E}tudes Spatiales (CNES).
MT is partially supported by the PRIN-INAF 2011 with the project
Transient Universe: from ESO Large to PESSTO.
This work was made possible in part by a research grant from the
Armenian National Science and Education Fund (ANSEF)
based in New York, USA.
Funding for SDSS-III has been provided by the Alfred P.~Sloan Foundation,
the Participating Institutions, the National Science Foundation,
and the US Department of Energy Office of Science.
The SDSS--III web site is \texttt{http://www.sdss3.org/}.
SDSS--III is managed by the Astrophysical Research Consortium for the
Participating Institutions of the SDSS--III Collaboration including the
University of Arizona, the Brazilian Participation Group,
Brookhaven National Laboratory, University of Cambridge,
University of Florida, the French Participation Group,
the German Participation Group, the Instituto de Astrofisica de Canarias,
the Michigan State/Notre Dame/JINA Participation Group,
Johns Hopkins University, Lawrence Berkeley National Laboratory,
Max Planck Institute for Astrophysics, New Mexico State University,
New York University, Ohio State University, Pennsylvania State University,
University of Portsmouth, Princeton University, the Spanish Participation Group,
University of Tokyo, University of Utah, Vanderbilt University,
University of Virginia, University of Washington, and Yale University.

\bibliography{snbibIV}

\label{lastpage}

\appendix
\section{SN AND HOST GALAXY DATA} 
\label{append}

\begin{table*} \centering
  \caption{List of \textit{arm} SNe.}
  \label{arm}
  \tabcolsep 5pt
  \begin{tabular}{llllllrr}
    \hline
  \multicolumn{1}{c}{SN} & \multicolumn{1}{c}{Type$^a$} & \multicolumn{1}{c}{Galaxy} & \multicolumn{1}{c}{Morph.$^a$} & \multicolumn{1}{c}{Arm Class} & \multicolumn{1}{c}{$R_{\rm SN}/R_{\rm 25}$} & \multicolumn{1}{c}{$d_1^{b}$} & \multicolumn{1}{c}{$d_2^{b}$} \\
	\multicolumn{1}{c}{(1)} & \multicolumn{1}{c}{(2)} & \multicolumn{1}{c}{(3)} & \multicolumn{1}{c}{(4)} & \multicolumn{1}{c}{(5)} & \multicolumn{1}{c}{(6)} & \multicolumn{1}{c}{(7)} & \multicolumn{1}{c}{(8)} \\
  \hline
1921B & II & NGC~3184 & Sc & NGD & 0.74 & 0.65: & 0.30: \\
1937F & II P: & NGC~3184 & Sc & NGD & 0.68 & 0.43 & 0.03 \\
1947A & II & NGC~3177 & Sb & NGD & 0.97 & 0.67 & $\cdots$ \\
1951H & II: & NGC~5457 & Sc & GD & 0.52 & 0.75 & 0.62 \\
1961U & II L & NGC~3938 & Sc & GD & 0.76 & 0.49 & 0.15 \\
1963J & Ia & NGC~3913 & Sc & GD & 0.17 & 0.44 & 0.09 \\
1964A & II pec & NGC~3631 & Sc & GD & 1.12 & 0.95 & 0.91 \\
1964L & Ic & NGC~3938 & Sc & GD & 0.19 & 0.16 & --0.70 \\
1965L & II P & NGC~3631 & Sc & GD & 0.67 & 0.58 & --0.10 \\
1966K & Ia: & MCG~+05--27--53 & Sab & GD & 0.72 & 0.50 & 0.08 \\
1967H & II: & NGC~4254 & Sc & GD & 0.47 & 0.39 & 0.02 \\
1970G & II L & NGC~5457 & Sc & GD & 0.55 & 0.83 & 0.73 \\
1972Q & II P & NGC~4254 & Sc & GD & 0.66 & 0.36 & --0.11 \\
1973U & II & IC~43 & SBc & NGD & 0.57 & 0.57 & 0.25 \\
1974J & Ia* & NGC~7343 & SBb & NGD & 0.57 & 0.05 & --0.73 \\
1975O & Ia* & NGC~2487 & SBc & GD & 0.45 & 0.49* & --0.02* \\
1978B & II & MCG~+10--16--117 & Sd & GD & 0.56 & 0.43 & --0.29 \\
1978H & II & NGC~3780 & Sc & NGD & 0.24 & 0.25 & --0.46 \\
1979C & II L & NGC~4321 & Sc & GD & 0.58 & 0.77 & 0.65 \\
1983I & Ib/c & NGC~4051 & SBbc & NGD & 0.40 & 0.25 & 0.02 \\
1986A & Ia & NGC~3367 & SBc & NGD & 0.29 & 0.63 & 0.46 \\
1986I & II P & NGC~4254 & Sc & GD & 0.23 & 0.29 & --0.29 \\
1987B & IIn L & NGC~5850 & SBb & GD & 1.11 & 0.73 & 0.47 \\
1988O & Ia & PGC~0054128 & Sc & NGD & 0.37 & 0.89 & 0.79 \\
1988R & Ia & MCG~+09--23--09 & Sc & NGD & 0.38 & 0.27* & --0.47* \\
1989K & II & NGC~5375 & SBbc & NGD & 0.68 & 0.66 & 0.29 \\
1990ag & II pec & A073249+3254 & Sc & GD & 0.34 & 0.48* & --0.04* \\
1991S & Ia & UGC~5691 & Sbc & GD & 0.65 & 0.44 & 0.06 \\
1992C & II & NGC~3367 & SBc & NGD & 0.48 & 0.51* & 0.02* \\
1992aa & IIn & NGC~6464 & SBc & GD & 0.61 & 0.46 & --0.11 \\
1992ad & II & NGC~4411B & Scd & NGD & 0.50 & 0.22 & --0.42 \\
1992am & II P & PGC~0005247 & SBb & GD & 0.81 & 0.75 & 0.57 \\
1993Z & Ia & NGC~2775 & Sa & NGD & 0.30 & 0.26 & --0.19 \\
1993ab & Ia & NGC~1164 & SBab & GD & 0.51 & 0.38 & 0.17 \\
1994A & II & UGC~8214 & SBb & GD & 0.35 & 0.02 & --0.95 \\
1994K & Ia & PGC~0028944 & SBbc & NGD & 0.81 & 0.77 & 0.54 \\
1995G & IIn & NGC~1643 & SBbc: & GD & 0.50 & 0.19 & --0.51 \\
1995Z & II & UGC~937 & SBd & NGD & 0.44 & 0.92 & 0.88 \\
1996an & II & NGC~1084 & Sc & NGD & 0.23 & 0.56* & 0.12* \\
1996cd & Ib/c: & A075720+1112 & SBbc & NGD & 0.53 & 0.25 & --0.37 \\
1997ef & Ic pec & UGC~4107 & Sc & NGD & 0.49 & 0.91* & 0.83* \\
1997ei & Ic & NGC~3963 & SBbc & GD & 0.18 & 0.37* & --0.25* \\
1998C & II & UGC~3825 & SBc & NGD & 0.35 & 0.02: & --0.95: \\
1998V & Ia & NGC~6627 & SBb & GD & 0.53 & 0.46 & --0.31 \\
1998ab & Ia pec & NGC~4704 & SBbc & NGD & 0.47 & 0.16* & --0.67* \\
1998aq & Ia & NGC~3982 & Sbc & NGD & 0.33 & 0.25 & --0.5 \\
1998dl & II & NGC~1084 & Sc & NGD & 0.24 & 1.00 & 1.00 \\
1999K & II & A080306+0324 & SBd & GD & 0.67 & 0.21 & $\cdots$ \\
1999bg & II: & IC~758 & SBd & GD & 0.61 & 0.43 & --0.14 \\
1999br & II pec & NGC~4900 & SBd & NGD & 0.57 & 0.79 & $\cdots$ \\
1999di & Ib & NGC~776 & Sbc & GD & 0.36 & 0.19: & --0.62: \\
1999dk & Ia & UGC~1087 & Sc & NGD & 0.65 & 0.71 & 0.41 \\
1999do & Ia & MCG~+05--54--03 & Sa: & NGD & 0.21 & 0.02 & --0.97 \\
1999ef & Ia & UGC~607 & SBc & GD & 0.69 & 0.64 & 0.25 \\
1999et & II & NGC~1643 & SBbc: & GD & 0.34 & 0.42 & --0.07 \\
1999gb & IIn & NGC~2532 & Scd & NGD & 0.31 & 0.51: & 0.03: \\
1999ge & II & NGC~309 & SBc & GD & 0.19 & 0.38 & --0.01 \\
1999gi & II P & NGC~3184 & Sc & NGD & 0.28 & 0.70: & 0.41: \\
1999gk & II & NGC~4653 & Scd & NGD & 0.61 & 0.55 & --0.28 \\
1999gn & II & NGC~4303 & SBbc & GD & 0.24 & 0.41* & --0.19* \\
2000F & Ic: & IC~302 & SBbc & NGD & 0.31 & 0.11 & $\cdots$ \\
2000J & II & UGC~8510 & SBc & GD & 0.82 & 0.09 & $\cdots$ \\
2000O & Ia & MCG~+03--31--61 & SBd & GD & 0.67 & 0.44 & 0.08 \\
  \hline \\
  \end{tabular}
 \end{table*}

\setcounter{table}{0} 

\begin{table*}\centering
\caption{\textit{Continued...}}
\tabcolsep 5pt
  \begin{tabular}{llllllrr}
	\hline
  \multicolumn{1}{c}{SN} & \multicolumn{1}{c}{Type$^a$} & \multicolumn{1}{c}{Galaxy} & \multicolumn{1}{c}{Morph.$^a$} & \multicolumn{1}{c}{Arm Class} & \multicolumn{1}{c}{$R_{\rm SN}/R_{\rm 25}$} & \multicolumn{1}{c}{$d_1^{b}$} & \multicolumn{1}{c}{$d_2^{b}$} \\
	\multicolumn{1}{c}{(1)} & \multicolumn{1}{c}{(2)} & \multicolumn{1}{c}{(3)} & \multicolumn{1}{c}{(4)} & \multicolumn{1}{c}{(5)} & \multicolumn{1}{c}{(6)} & \multicolumn{1}{c}{(7)} & \multicolumn{1}{c}{(8)} \\
  \hline
2000cq & II & UGC~10354 & Scd & GD & 0.88 & 0.29 & --0.55 \\
2000ct & IIn & A170103+3328 & SBd & NGD & 0.66 & 0.72 & $\cdots$ \\
2000dq & II & MCG~+00--06--43 & Sbc & GD & 0.55 & 0.09* & --0.82* \\
2000du & II & UGC~3920 & Sb & GD & 0.42 & 0.60 & 0.24 \\
2001J & II & UGC~4729 & SBd & NGD & 0.55 & 0.78 & 0.17 \\
2001Q & II & UGC~6429 & Sc & GD & 0.42 & 0.53 & --0.06 \\
2001Y & II pec & NGC~3362 & Sc & GD & 0.46 & 0.50 & --0.14 \\
2001aj & II & UGC~10243 & Sc & NGD & 0.52 & 0.91 & 0.84 \\
2001fv & II & NGC~3512 & Sc & NGD & 0.54 & 0.27: & --0.47: \\
2001it & II & MCG~+09--25--15 & Sc & GD & 0.33 & 0.29 & --0.24 \\
2002ap & Ic pec & NGC~628 & Sc & GD & 0.91 & 0.88 & 0.75 \\
2002at & II: & NGC~3720 & Sb & NGD & 0.22 & 0.50* & 0.00* \\
2002ca & II & UGC~8521 & SBab & NGD & 0.41 & 0.97: & 0.94: \\
2002cb & IIn & MCG~+08--24--34 & SBc & GD & 0.56 & 0.43 & 0.14 \\
2002cp & Ib/c & NGC~3074 & Scd & NGD & 0.93 & 0.64 & --0.14 \\
2002df & Ia & MCG~--01--53--06 & SBc & NGD & 0.72 & 0.02 & --0.97 \\
2002en & II & UGC~12289 & SBc & GD & 0.41 & 0.28 & --0.30 \\
2002fj & IIn & NGC~2642 & SBbc & GD & 0.32 & 0.44: & --0.12: \\
2002ji & Ib/c & NGC~3655 & Sbc & NGD & 0.35 & 0.14 & --0.72 \\
2003C & II & UGC~439 & Sb & GD & 0.22 & 0.69 & 0.59 \\
2003G & IIn & IC~208 & Sc & NGD & 0.23 & 0.05 & --0.90 \\
2003L & Ic & NGC~3506 & Sc & NGD & 0.23 & 0.36 & --0.06 \\
2003T & II & UGC~4864 & Sb & GD & 0.52 & 0.25 & --0.54 \\
2003cq & Ia & NGC~3978 & Sc & NGD & 0.52 & 0.25 & --0.54 \\
2003ej & II & UGC~7820 & Scd & NGD & 0.64 & 0.53 & --0.07 \\
2003el & Ic & NGC~5000 & SBbc & GD & 0.32 & 0.51: & 0.01: \\
2003fd & Ia & UGC~8670 & SBcd & GD & 0.61 & 0.55 & 0.25 \\
2003gd & II P & NGC~628 & Sc & GD & 0.62 & 0.64 & 0.30 \\
2003hq & Ia & MCG~+08--30--27 & SBbc & GD & 0.38 & 1.00 & 1.00 \\
2003ih & Ib/c & UGC~2836 & Sa & NGD & 0.16 & 0.77 & 0.31 \\
2003iy & II: & NGC~6143 & Sc & NGD & 0.21 & 0.13 & --0.57 \\
2003kw & II & UGC~6314 & SBbc & GD & 0.30 & 0.31 & --0.53 \\
2004K & Ia & E579--G22 & SBb & GD & 0.51 & 0.03 & --0.94 \\
2004T & II & UGC~6038 & Sb & GD & 0.38 & 0.55 & 0.05 \\
2004bk & Ia & NGC~5246 & SBb & GD & 0.36 & 0.48* & --0.03* \\
2004ef & Ia & UGC~12158 & SBbc & GD & 0.26 & 0.27 & --0.47 \\
2004er & II & MCG~--01--07--24 & Scd & NGD & 0.63 & 0.71 & 0.41 \\
2004es & II & UGC~3825 & SBc & NGD & 0.69 & 0.50 & --0.16 \\
2004ey & Ia & UGC~11816 & SBc & NGD & 0.28 & 0.27 & --0.36 \\
2004gx & II: & UGC~12663 & SBc & NGD & 0.29 & 0.69 & 0.54 \\
2005I & II & IC~983 & SBbc & NGD & 0.35 & 0.69 & 0.19 \\
2005ao & Ia & NGC~6462 & Sb & NGD & 0.65 & 0.33 & --0.38 \\
2005ay & II P & NGC~3938 & Sc & GD & 0.35 & 0.16 & --0.45 \\
2005bh & Ic & UGC~6495 & SBc & GD & 0.53 & 0.40 & --0.15 \\
2005bj & Ic: & MCG~+03--43--05 & SBd & NGD & 0.47 & 0.63 & 0.00 \\
2005bk & Ic & MCG~+07--33--27 & Sab & NGD & 0.19 & 0.06 & --0.86 \\
2005cl & IIn & MCG~--01--53--20 & SBb & GD & 0.47 & 0.02 & --0.91 \\
2005cv & II & UGC~1359 & Sb & GD & 0.22 & 0.19 & --0.36 \\
2005dz & II & UGC~12717 & Sc & NGD & 0.66 & 0.26 & --0.49 \\
2005es & II & MCG~+01--59--79 & Sbc & GD & 0.34 & 0.25 & --0.14 \\
2005gl & IIn & NGC~266 & SBb & GD & 0.28 & 0.46: & --0.07: \\
2005gm & II & NGC~1423 & SBb & GD & 0.43 & 0.44 & 0.06 \\
2005lt & Ia & MCG~+03--30--51 & SBc & NGD & 0.62 & 0.80 & 0.53 \\
2006G & IIb & NGC~521 & SBbc & GD & 0.47 & 0.50 & --0.06 \\
2006ac & Ia & NGC~4619 & SBbc & NGD & 0.44 & 0.94 & 0.92 \\
2006ad & II & A090743+1203 & SBd & GD & 0.81 & 0.50 & $\cdots$ \\
2006bl & II & MCG~+02--40--09 & Sc & NGD & 0.32 & 0.50 & 0.03 \\
2006cp & Ia & UGC~7357 & SBd & NGD & 0.53 & 0.69 & 0.46 \\
2006cy & IIn & A130801+2606 & Sc & NGD & 0.54 & 0.30 & --0.54 \\
2006dl & IIb & MCG~+04--31--05 & Sc: & NGD & 0.40 & 0.07 & --0.81 \\
2006dp & II & MCG~--01--03--56 & Sbc & GD & 0.23 & 0.06 & --0.88 \\
2006en & Ia & MCG~+05--54--41 & Sbc: & NGD & 0.19 & 0.11* & --0.79* \\
2006es & Ia & UGC~2828 & SBbc & GD & 0.28 & 0.83: & 0.65: \\
\hline \\
  \end{tabular}
\setcounter{table}{0} 
\end{table*}

\setcounter{table}{0} 

\begin{table*}\centering
\caption{\textit{Continued...}}
\tabcolsep 5pt
  \begin{tabular}{llllllrr}
	\hline
  \multicolumn{1}{c}{SN} & \multicolumn{1}{c}{Type$^a$} & \multicolumn{1}{c}{Galaxy} & \multicolumn{1}{c}{Morph.$^a$} & \multicolumn{1}{c}{Arm Class} & \multicolumn{1}{c}{$R_{\rm SN}/R_{\rm 25}$} & \multicolumn{1}{c}{$d_1^{b}$} & \multicolumn{1}{c}{$d_2^{b}$} \\
	\multicolumn{1}{c}{(1)} & \multicolumn{1}{c}{(2)} & \multicolumn{1}{c}{(3)} & \multicolumn{1}{c}{(4)} & \multicolumn{1}{c}{(5)} & \multicolumn{1}{c}{(6)} & \multicolumn{1}{c}{(7)} & \multicolumn{1}{c}{(8)} \\
  \hline	
2006gs & II & NGC~3977 & Sc & NGD & 0.21 & 0.45 & --0.20 \\
2006jd & IIn & UGC~4179 & SBbc & NGD & 0.57 & 0.44* & --0.13* \\
2007F & Ia & UGC~8162 & SBc & GD & 0.36 & 0.13 & $\cdots$ \\
2007K & IIn & MCG~+06--20--50 & SBb & GD & 0.30 & 0.17 & --0.71 \\
2007O & Ia & UGC~9612 & SBc & GD & 0.19 & 0.67 & 0.26 \\	
2007aa & II & NGC~4030 & Sbc & GD & 0.67 & 0.74 & 0.46 \\
2007ad & II & UGC~10845 & Sc & NGD & 0.31 & 0.42 & --0.15 \\
2007am & II & NGC~3367 & SBc & NGD & 0.25 & 0.60 & 0.31 \\
2007aq & II P & IC~2409 & Sbc & NGD & 0.90 & 0.53 & --0.15 \\
2007ay & IIb & UGC~4310 & Scd & NGD & 0.48 & 0.26 & --0.17 \\
2007bc & Ia & UGC~6332 & Sab & GD & 0.78 & 0.48 & $\cdots$ \\
2007bp & II & A092400+0512 & Sc & GD & 0.36 & 0.18 & --0.62 \\
2007bv & II & A153410+0705 & SBb & GD & 0.39 & 0.34 & 0.13 \\
2007cl & Ic & NGC~6479 & Sc & GD & 0.26 & 0.22 & --0.14 \\
2007gl & Ib/c & A031133--0044 & Sbc & GD & 0.54 & 0.62 & 0.26 \\
2007hb & Ib/c & NGC~819 & Sbc & GD & 0.45 & 0.84 & 0.77 \\
2007il & II & IC~1704 & Sc & NGD & 0.43 & 0.59 & 0.19 \\
2007pk & IIn pec & NGC~579 & Scd & NGD & 0.15 & 0.59 & 0.20 \\
2007rt & IIn & UGC~6109 & Sc & GD & 0.15 & 0.23 & --0.53 \\
2008ae & Ia pec & IC~577 & Sbc & GD & 0.55 & 0.88 & 0.80 \\
2008bh & II & NGC~2642 & SBbc & GD & 0.38 & 0.74 & 0.58 \\
2008bl & II & UGC~9317 & SBcd & NGD & 0.32 & 0.35 & --0.28 \\
2008bs & Ib & UGC~4085 & Sc & NGD & 0.22 & 0.33 & --0.20 \\
2008cx & IIb & NGC~309 & SBc & GD & 0.63 & 0.65 & 0.07 \\
2008du & Ic & NGC~7422 & SBb & NGD & 0.24 & 0.45 & --0.08 \\
2008dz & II & NGC~5123 & Sc & GD & 0.68 & 0.65 & 0.37 \\
2008ew & Ic & IC~1236 & SBc & GD & 0.33 & 0.11 & --0.75 \\
2008ie & IIb & NGC~1070 & Sc & NGD & 0.26 & 0.00 & --1.00 \\
2008in & II P & NGC~4303 & SBbc & GD & 0.50 & 0.92 & 0.87 \\
2009W & II P & A162347+1144 & SBb & NGD & 1.14 & 0.42 & --0.03 \\
2009af & II & UGC~1551 & SBd & NGD & 0.26 & 0.06 & --0.84 \\
2009ay & II & NGC~6479 & Sc & GD & 0.34 & 0.23* & --0.54* \\
2009bv & Ia & MCG~+06--29--39 & Sc & GD & 0.73 & 0.73 & 0.51 \\
2009cz & Ia & NGC~2789 & Sa & NGD & 0.38 & 0.20 & --0.33 \\
2009en & Ia & UGC~9515 & SBc & NGD & 0.57 & 0.12: & --0.76: \\
2009gk & IIb & UGC~11803 & Sbc & NGD & 0.70 & 0.34 & 0.05 \\
2009ha & Ib & MCG~--01--07--24 & Scd & NGD & 0.40 & 0.43 & --0.03 \\
2009ik & Ia & NGC~4653 & Scd & NGD & 0.97 & 0.67 & --0.05 \\
2009ls & II & NGC~3423 & Scd & NGD & 0.34 & 0.15 & --0.81 \\
2009lx & II P & MCG~+01--30--08 & SBb & GD & 0.77 & 0.34 & --0.33 \\
2010av & Ib/c: & IC~1099 & SBc & NGD & 0.54 & 0.69 & 0.39 \\
2010bs & II & UGC~7416 & SBb & GD & 0.70 & 0.55 & --0.07 \\
2010ct & II P: & NGC~3362 & Sc & GD & 0.72 & 0.44 & --0.17 \\
2010fz & Ia & NGC~2967 & Sc & NGD & 0.27 & 0.83 & 0.63 \\
2010gs & II & MCG~--01--53--07 & Sbc & GD & 0.65 & 0.52 & 0.08 \\
2010ig & Ib & UGC~1306 & SBa & NGD & 0.83 & 0.53 & 0.34 \\
2010jc & II P & NGC~1033 & Sc & NGD & 0.90 & 0.23 & $\cdots$ \\
2011ak & II P & UGC~6997 & Sd & GD & 0.54 & 0.43 & --0.29 \\
2011an & IIn & UGC~4139 & SBcd & NGD & 0.41 & 0.52 & 0.06 \\
2011bc & Ia & NGC~4076 & Sbc & NGD & 0.29 & 1.00 & 1.00 \\
2011bi & II P & MCG~+07--35--37 & Sc & GD & 0.34 & 0.33 & --0.50 \\
2011bk & Ia & A162034+2112 & SBbc & GD & 0.31 & 0.60 & 0.18 \\
\hline \\
  \end{tabular}
  \parbox{\hsize}{$^a$ Among all the SNe and host galaxy morphological types,
    there are some uncertain 
                  (`:' or `?') and peculiar (`pec') classifications.
                  In two cases, marked by `*', the Ia~type has been inferred from the light curve only.\\
                  $^b$ The distance is flagged by a `*' or `:' symbol
									when one of the edges of the spiral arm is roughly determined.
									In these cases, the contaminated (by the light of a star or by the SN itself)
									edge is determined as a symmetric reflection
									of the other edge from the peak.
									The `*' symbol indicates that the SN
                  is located between the peak and the roughly determined edge of the arm,
									while the `:' symbol indicates that the SN is located between the well-determined edge and the peak.
									In 10 cases, the $d_2$ parameter was not determined because of a noisy arm profile.}
\setcounter{table}{0} 
\end{table*}

\setcounter{table}{1} 

\begin{table*} \centering
  \caption{List of \emph{interarm} SNe.}
  \label{interarm}
  \tabcolsep 5pt
  \begin{tabular}{llllllrr}
    \hline
  \multicolumn{1}{c}{SN} & \multicolumn{1}{c}{Type$^a$} & \multicolumn{1}{c}{Galaxy} & \multicolumn{1}{c}{Morph.$^a$} & \multicolumn{1}{c}{Arm Class} & \multicolumn{1}{c}{$R_{\rm SN}/R_{\rm 25}$} & \multicolumn{1}{c}{$d_1^{b}$} & \multicolumn{1}{c}{$d_2^{b}$} \\
	\multicolumn{1}{c}{(1)} & \multicolumn{1}{c}{(2)} & \multicolumn{1}{c}{(3)} & \multicolumn{1}{c}{(4)} & \multicolumn{1}{c}{(5)} & \multicolumn{1}{c}{(6)} & \multicolumn{1}{c}{(7)} & \multicolumn{1}{c}{(8)} \\
  \hline
    1941C & II & NGC~4136 & SBc & NGD & 0.68 & --0.37 & $\cdots$ \\
		1963D & Ia* & NGC~4146 & SBb & GD & 0.65 & 1.09 & 1.18 \\
		1963P & Ia & NGC~1084 & Sc & NGD & 0.26 & 1.12 & 1.23 \\
		1971U & Ia & MCG~+05--26--14 & SBcd & NGD & 0.25 & --0.22 & --1.37 \\
		1978E & Ia & MCG~+06--49--36 & Sc & NGD & 0.53 & --0.19 & --1.39 \\
		1979B & Ia & NGC~3913 & Sc & GD & 0.51 & --0.03 & $\cdots$ \\
		1987C & IIn P & MCG~+09--14--47 & SBc & NGD & 0.72 & 1.05 & 1.08 \\
		1989A & Ia & NGC~3687 & SBc & NGD & 0.44 & 1.71 & 2.41 \\
		1991am & Ia & MCG~+06--37--06 & Sb & NGD & 0.30 & --0.56 & --2.30 \\
		1992ap & Ia & UGC~10430 & SBbc & NGD & 0.62 & 1.20 & 1.34 \\
		1997aa & II & IC~2102 & SBd & NGD & 0.85 & --0.06 & $\cdots$ \\
		2001fa & IIn & NGC~673 & Sc & GD & 0.27 & 1.03 & $\cdots$ \\
		2001gb & Ia & IC~582 & Sc & NGD & 0.51 & --0.15 & $\cdots$ \\
		2002hw & Ia & UGC~52 & Sc & NGD & 0.16 & 1.92 & 3.30 \\
		2003P & Ia & MCG~+09--13--107 & Sbc & NGD & 0.23 & --0.42 & $\cdots$ \\
		2003U & Ia & NGC~6365A & Sc & GD & 0.35 & 1.17 & 1.30 \\
		2003ie & II & NGC~4051 & SBbc & NGD & 0.63 & 1.07 & $\cdots$ \\
		2004G & II & NGC~5668 & Sd & NGD & 0.41 & --0.05 & $\cdots$ \\
		2004dv & II & MCG~--01--06--12 & SBcd & NGD & 0.83 & 1.18 & 1.30 \\
		2005dq & II & UGC~12177 & SBc & GD & 0.20 & --0.06 & $\cdots$ \\
		2005he & II & MCG~+06--49--68 & SBc & NGD & 0.23 & --0.29 & --1.58 \\
		2005ms & Ia & UGC~4614 & Scd & NGD & 1.16 & 1.40 & $\cdots$ \\
		2006V & II & UGC~6510 & SBcd & GD & 0.77 & --0.15 & $\cdots$ \\
		2006X & Ia & NGC~4321 & Sc & GD & 0.29 & --0.11 & $\cdots$ \\
		2006dh & Ia & UGC~8670 & SBcd & GD & 0.39 & 1.79 & 2.74 \\
		2007kk & Ia & UGC~2828 & SBbc & GD & 0.37 & --0.66 & --2.32 \\
		2008dr & Ia & NGC~7222 & SBbc & GD & 0.23 & --1.46 & --3.9 \\
		2008el & II P & PGC~70567 & Sbc & GD & 1.00 & 1.17 & $\cdots$ \\
		2008fe & II P & UGC~9578 & SBbc & NGD & 0.75 & --0.07 & $\cdots$ \\
		2008gi & II & A024400+0525 & SBd & NGD & 0.17 & --0.15 & --1.55 \\
		2009I & Ia & NGC~1080 & Sc & NGD & 0.35 & --0.05 & --1.10 \\
		2009ig & Ia & NGC~1015 & SBa & NGD & 0.30 & --0.01 & --1.02 \\
		2009iq & II P & UGC~2308 & Sc & GD & 0.25 & --0.11 & --1.21 \\
		2010B & Ia & NGC~5370 & SBa & NGD & 0.23 & --0.41 & --2.60 \\
		2010Z & II & NGC~2797 & Sb & GD & 0.29 & --0.08 & --1.12 \\
		2010hz & Ia & UGC~1359 & Sb & GD & 0.20 & 1.38 & 1.77 \\
		2010ii & Ia & NGC~7342 & SBab & GD & 0.64 & --0.27 & --1.41 \\
	\hline \\
  \end{tabular}
  \parbox{\hsize}{The symbols used in this table are the same as in Table~\ref{arm}.}
\end{table*}

\end{document}